# A Search for Laser Emission with Megawatt Thresholds from 5600 FGKM Stars

Nathaniel K. Tellis[1], and Geoffrey W. Marcy[2]

[1]Astronomy Department, University of California, Berkeley, CA, 94720 USA. Nate.tellis@gmail.com. (Corresponding author)

[2]Astronomy Department, University of California, Berkeley, CA, 94720 USA. Professor Emeritus

Abstract

We searched high resolution spectra of 5600 nearby stars for emission lines that are both inconsistent with a natural origin and unresolved spatially, as would be expected from extraterrestrial optical lasers. The spectra were obtained with the Keck 10-meter telescope, including light coming from within 0.5 arcsec of the star, corresponding typically to within a few to tens of au of the star, and covering nearly the entire visible wavelength range from 3640 to 7890 Å. We establish detection thresholds by injecting synthetic laser emission lines into our spectra and blindly analyzing them for detections. We compute flux density detection thresholds for all wavelengths and spectral types sampled. Our detection thresholds for the power of the lasers themselves range from 3 kW to 13 MW, independent of distance to the star but dependent on the competing "glare" of the spectral energy distribution of the star and on the wavelength of the laser light, launched from a benchmark, diffraction-limited 10-meter class telescope. We found no such laser emission coming from the planetary region around any of the 5600 stars. As they contain roughly 2000 lukewarm, Earth-size planets, we rule out models of the Milky Way in which over 0.1% of warm, Earth-size planets harbor technological civilizations that, intentionally or not, are beaming optical lasers toward us. A next generation spectroscopic laser search will be done by the *Breakthrough Listen* initiative, targeting more stars, especially stellar types overlooked here including spectral types O, B, A, early F, late M, and brown dwarfs, and astrophysical exotica.

1. Introduction

Since the 1950's, the Search for extraterrestrial intelligence (SETI) has involved monitoring many electromagnetic wavelength bands for signals of non-astrophysical origin (e.g. Cocconi & Morrison 1959; Drake 1961; Horowitz & Sagan (1993); Tarter 2001; Howard et al. 2007; Hanna et al. 2009; Siemion et al. 2013; Wright et al. 2014a; Wright et al. 2016; Lacki 2016)). Searches have been conducted at radio wavelengths with Arecibo, Green Bank telescope, Parkes radio telescope, and the Allen telescope array (Drake 1961; Werthimer et al. 2001; Korpela et al. 2011; Siemion et al. 2013, Tarter et al. 2011, and Harp et al. 2015). These searches targeted both stars and galaxies, and more recently stars harboring known exoplanets (Siemion et al. 2013). Searches for both continuously transmitting and short duration light pulses in the visible spectrum (optical SETI) have been conducted (Wright et al. 2001; Reines and Marcy 2002; Howard et al. 2004; Stone et al. 2005; Howard et al. 2007, Hanna et al. 2009, Tellis & Marcy 2015, Abeysekara et al. 2016). A unique search for near-infrared pulses has also begun (Wright et al. 2014c, Maire et al. 2014). At the time of writing, none of these searches has yielded any convincing evidence of other technological civilizations.

Searches are also being conducted for extraterrestrial artifacts, detected either by the waste thermal emission from large machinery sometimes referred to as "Dyson spheres" (Dyson 1960; Wright et al. 2015), or by simple occlusion of starlight by a non-natural intervening object (Arnold 2005, Forgan 2013, Walkowicz et al. 2014). This latter technique was put to the test recently with the strange photometry of KIC 8462852, commonly referred to as Boyajian's Star. This star has been since imaged in the visible, infrared, and radio domains, yielding prospective explanations for its unusual light curve that include breakup of comets near the star, unusual inhomogeneities and pulsations on the stellar surface, passing dust clouds, and consumption of an orbiting planet (Schuetz et al. 2015, Marengo et al. 2015, Abeysekara et al. 2016, Bodman & Quillen 2016, Boyajian et al. 2016, Wright et al. 2016, Metzger et al. 2017).

There are numerous technologically innovative SETI projects being launched and planned. The *Breakthrough Listen* initiative began taking data in January 2016 and includes ambitious radio-wave searches with novel electronics to gather data in bandwidths of up to 4 GHz with the Green Bank and Parkes radio telescopes. *Breakthrough Listen* has also begun a spectroscopic search for optical laser emission on the 2.4-meter Automated Planet Finder of Lick Observatory (see Section 7). Commensal, beam-forming radio observations with the future MeerKAT array and the Square Kilometer Array, (Siemion et al. 2014) would be 100x more sensitive to SETI radio signals than prior searches and would have the ability to survey millions of stars and galaxies limited by back-end compute power. The interest in optical and near-infrared extraterrestrial laser emission has spurred searches for pulses having duration as short as a nanosecond or displaying periodicities (Howard et al. 2004, 2007; Korpela et al. 2011; Drake et al. 2010; Mead 2013; Covault 2013; Leeb et al. 2013; Gillon 2014; Wright et al. 2014ab). Discussions are proceeding about carrying out all-sky searches for optical pulses. There is also renewed interest in detecting the light from propulsion systems based on radiation pressure for interstellar spacecraft, such as the near-IR Gigawatt laser system outlined by the *Breakthrough StarShot* initiative (Zubrin 1995, Lubin et al. 2016, Kipping & Teachey 2016, http://breakthroughinitiatives.org/Initiative/3).

Several factors favor detection of extraterrestrial intelligence by their optical or IR laser emission (Townes & Schwartz 1961; Zuckerman 1985). A high-intensity, diffraction-limited optical laser beam could transmit data at a high rate (exceeding $10^{12}$ bits per second) over Galactic distances, while keeping beam opening angles narrow, well under an arcsecond, minimizing leakage and eavesdropping (Townes & Schwartz 1961; Wright et al. 2001, Howard et al. 2007). The pencil beams afforded by Optical and IR radiation allow for low relative transmission powers and privacy. Thus, interstellar communication by UV, optical or IR lasers may be preferred over longer wavelengths. Additionally, the monochromatic nature of lasers used for communication allows them to outshine their host star in their small wavelength band. Laser lines as narrow as 1 Hz are already in use on Earth. The combination of beam angles under 1 arcsec and narrow wavelength bands allows modern-day lasers of mere megawatt power to have a higher specific intensity than that of the Sun, and much higher than that of the most common stars, the K- and M-type main sequence stars. For reference, continuous wave 150 kilowatt lasers have been accepted for field testing (DARPA 2015), and several pulsed lasers generate Petawatt power levels albeit for durations of picoseconds.

In this paper we present the results of an inspection of a large set of legacy echelle spectroscopic observations for monochromatic emission that could be consistent with a laser of extraterrestrial origin.

## 2. Target Stars and Spectroscopic Observations
### 2.1 Target Stars

The 67,708 Keck-HIRES spectra included in this survey include all spectra taken as part of the California Planet Search (CPS) from August 2004, when the HIRES spectrometer was upgraded to three CCDs, to March 2016. For a technical description of the HIRES spectrometer, see (Vogt et al., 1994). There are a total of 5600 target stars, most being main sequence and subgiant stars of spectral type FGKM. They span the full range of RA and DEC visible from Hawaii, as shown in Figure 1. They span ages from under 1 billion years (a few hundred stars are younger than 200 Myr, including some stars in young open clusters and moving groups), to stars age-dated (typically by chromospheric emission) at nearly 10 billion years (Marcy et al. 2008, Howard et al. 2012, Wright et al. 2011, Isaacson et al. 2017).

This set includes target stars in the CPS, a program that continues to make repeated Doppler measurements of over 3000 FGKM main sequence stars typically brighter than Vmag = 12 (most being brighter than Vmag=9) and northward of declination -25 deg. Numerous papers have been written about the target stars (e.g., Marcy 2008; Johnson et al. 2011; Howard et al. 2014). Roughly 10% of these CPS targets have known planets as detected by the Radial Velocity method. The search has been ongoing for the past 10-20 years, using both the Lick and Keck Observatories, though Keck 1 is now primarily used since the retirement of the iodine cell at the Lick Observatory 3-m Shane telescope (Fischer et al. 2014; Howard et al. 2014; Marcy et al. 2008; Wright et al. 2011).

Our target stars include spectra taken with Keck HIRES of all 1100 *Kepler* Objects of Interest (KOI) brighter than Kepmag=14.2, identified before the end of 2014, as well as some spectra of another 350 fainter KOIs harboring multiple transiting planets. This *Kepler* set mostly consists of main sequence

FGK and M-type stars, with a distribution of effective temperatures similar to that of the CPS targets shown in Figure 2. Over 90% of these *Kepler* objects of interest (KOI) are believed to harbor planets despite not being conclusively confirmed (Morton & Johnson 2011; Morton et al. 2016). A list of these KOIs and measurements of each star's effective temperature (Teff), surface gravity (log g), iron abundance ([Fe/H], and projected rotational velocity (Vsini) are provided by (Petigura 2015; Howard et al. 2015).

We also include analysis here of Keck-HIRES spectra taken by the CPS team of over 200 stars that were identified in other planet-transit surveys as harboring at least one transiting planet, namely, TrES, HAT, WASP, XO, COROT (Alonso et al. 2004, Bakos et al. 2012, Bakos et al. 2013, Collier et al. 2007, Enoch et al. 2011, Cameron et al. 2007, McCullough et al. 2006, Moutou et al. 2013). For the HAT planet-transit survey, we also analyzed for laser emission the spectra of many candidate planet-transit stars, designated with their "HTR" numbers, even if the planet turned out to be a false positive.

A few dozen of the spectra analyzed here were not of stars but of galaxies, nebulae, supernovae, and other exotica observed as part of other research programs. Nonetheless, these spectra are amenable to the same search for laser emission as the FGKM stars, though our sensitivity to laser emission from these targets varies greatly on a case-by-case basis.

We provide an online table of all 67,708 spectra analyzed here, listing the target names, the UT date and time of the exposure, the exposure duration, and whether the iodine cell was in or not. The reader is referred to the Keck Observatory Archive to obtain the original spectra, raw and reduced if desired, which contain complete FITS headers with details of each exposure including the equatorial coordinates of the telescope (with flexure), the hour angle, and many parameters of both the telescope and the spectrometer. Most of the targets have names that can be identified successfully with the SIMBAD search engine. However, some stars have internal names that do not appear on SIMBAD, such as internal project names of candidate planets identified in transiting planet surveys. For example, stars with names starting with "HTR" are candidates identified in the HAT transit planet search, analogous to the Kepler Objects of Interest (KOI) in the NASA Kepler transiting planet search. For such objects, the interested reader can cross-identify our name in the Keck Observatory Archive to determine its identity.

Figure 1 shows the location in celestial coordinates of all 5600 targets in this search for laser emission. The targets are located at the full range of RA and mostly north of Declination -35 deg, constrained by the northern +19.7 deg latitude of the Keck Observatory. The targets are distributed roughly uniformly in RA, befitting the smooth distribution of stars within ~100 pc of the Sun in the disk of the Milky Way Galaxy. Superimposed on that distribution is the dense set of 2132 *Kepler* target stars observed within the NASA-*Kepler* field at RA = 18h 40m to 20h 08m and Dec = +37 to +52 deg. Most of these stars were observed as part of the *Kepler* Follow-up Observation Program (KFOP) to characterize the planets and host stars of the candidate planets, the vast majority of which do exist (Morton et al. 2016). Figure 1 also reveals patches of targets faintly apparent from upper left to lower right, representing the targets of the NASA K2 program, still being carried out by the *Kepler* telescope. Thus, this search for laser lines constitutes a broad selection of nearby stars within 100 pc, systems with transiting planets within 1500 pc, and over 400 known multi-planet systems, all constituting a broad distribution of pencil beams piercing both the Solar System (including the Ecliptic) and the solar neighborhood of the Milky Way Galaxy.

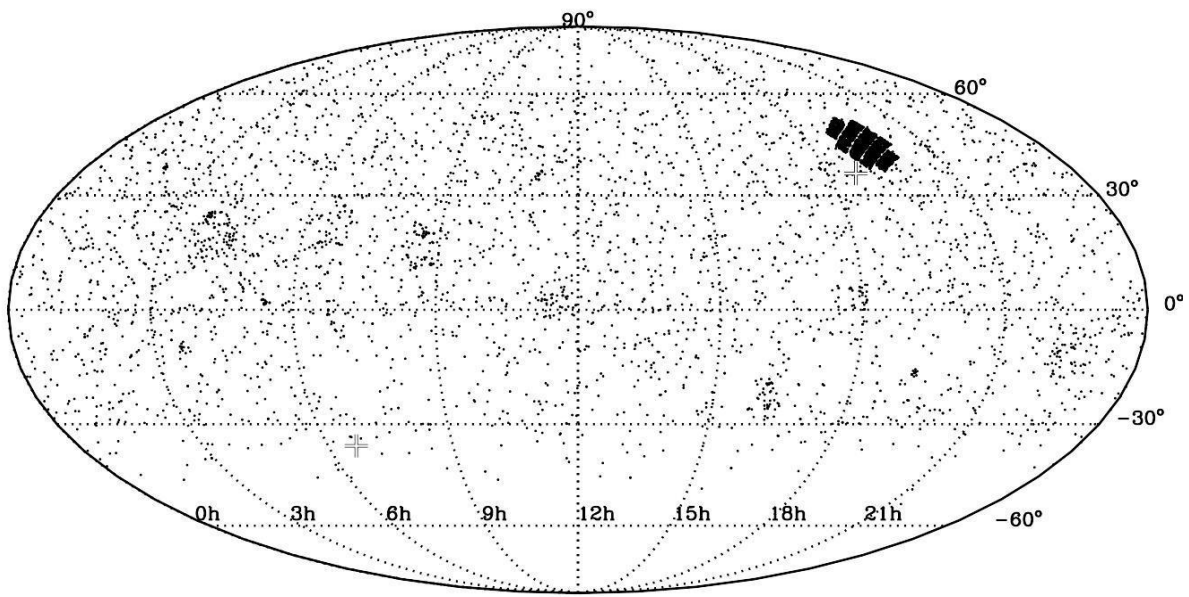

Fig 1. Positions in celestial coordinates of the 5600 targets in this search for laser emission lines. The targets consist mostly of nearby (within 100 pc) FGKM-type stars distributed roughly uniformly over the sky accessible to the Keck Observatory. Also prominent are the stellar targets at distances of 100-1000 pc observed within the field of view of the NASA-*Kepler* telescope at RA=19-20 hr and Dec=+40-50, and its faint footprints of fields of view of the NASA K2 mission along the ecliptic. This search for laser lines constitutes a broad selection of nearby stars and transiting planets.

Figure 2 shows a histogram of the effective temperatures of the 2600 nearby target stars observed in the CPS program, which are representative of the vast majority of the 5600 stars studied here (including the population observed as part of *Kepler* follow-up), most having been observed for precise Doppler measurements.

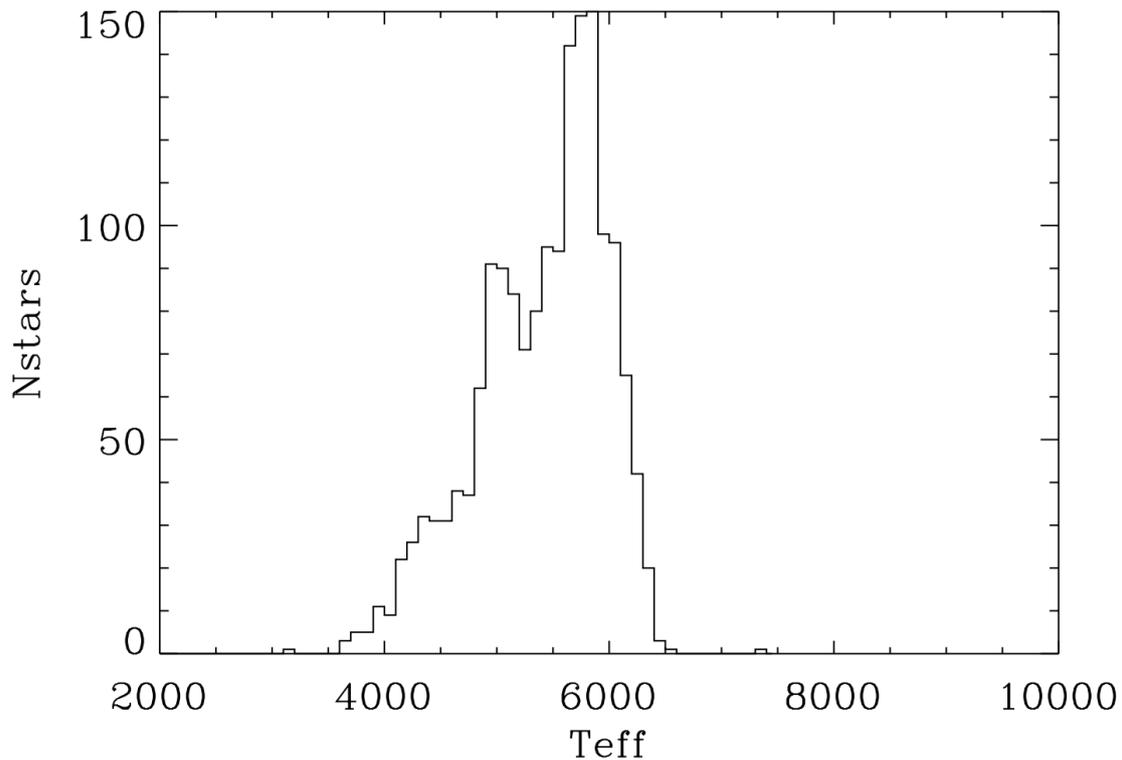

Fig 2. Histogram of effective temperature, Teff, of the 2800 FGKM stars observed in the California Planet Search program, typically including 5-50 exposures for each star to detect Doppler variations. The distribution of Teff spans the range 3500 to 6500K corresponding to early M-type stars to late F-type stars, in turn corresponding to stars with masses of 0.5 to 1.5 solar masses. This distribution is representative of the remaining 3300 stars, most of which were also observed for Doppler measurements to study exoplanets for which this range of Teff is required. This search for laser emission was confined to this narrow range of stellar types.

These spectra permit a search for pulsed and continuous laser emission, as long as the integrated power during our exposure is above our threshold, and the flux density is high enough over a small wavelength band (Section 5). We have previously carried out two similar searches for laser lines from subsets of these spectra, both looking for laser emission that stands above the stellar spectrum (and its noise) and for laser emission coming from the dark sky located 3 to 7 arcseconds away from the star (Reines & Marcy 2002, Tellis & Marcy 2015). This present search constitutes an advance over those past two spectroscopic efforts, by including over twice the wavelength coverage, an order of magnitude more stars, and in searching within ~0.5 arcsec of the stars, a tight field of view encompassing any planets or machines in the inner few au, including the habitable zones, of all target stars.

## 2.2 Spectroscopic Observations

The spectra analyzed here for laser emission were all obtained with the Keck-HIRES spectrometer on the Keck 1 telescope (Vogt et al. 1994). The spectra include nearly the entire wavelength range from 3640 to 7890 Å, with only small gaps in the spectra in between the spectral orders longward of 5600 Å. The spectral resolution for all spectra was $\lambda/\Delta\lambda = 60,000$, with the FWHM of the instrumental profile, $\Delta\lambda$, spanning 4.3 pixels in the wavelength direction and a Doppler width of 5.6 km s$^{-1}$, with variations of roughly 10% in that resolution as a function of wavelength. The pixels of the Keck-HIRES spectrometer span approximately 0.021 Å at a wavelength of 5500 Å and more generally span a wavelength range of $\lambda/250000$.

Thus, emission from a monochromatic laser captured by the Keck telescope would appear as an emission line of width 4.3 pixels in the wavelength direction, and would appear as a "dot" in space having a defined width in the spatial direction due to refraction of the light passing through the inhomogeneities in the Earth's atmosphere ("seeing"). The laser lines will be Doppler shifted by different amounts over time scales of hours, days, and months, due to any acceleration of the laser emitter and of the telescope due to orbital and rotational motion of the Earth. During a given typical ten-minute exposure, the laser emission line will be Doppler smeared due mostly to the Earth's rotation and (to a lesser extent) orbital motion by no more than a few percent of the instrumental profile width (FWHM=5.6 km s$^{-1}$), which does not significantly compromise our search algorithm for the emission lines. The CCD pixels were binned on chip by three pixels (1.1 arcsec) in the spatial dimension to reduce readout time. This spatial binning still allows us to distinguish between laser emission that comes from a spatially unresolved dot (as expected for emission from beyond the solar system) and spatially extended emission (such as from nebulosity or local scattering of light in the Earth's atmosphere). Hereafter, "pixels" will refer to the on-chip binned sets of three pixels.

All exposure times were normally limited by an exposure meter inside the HIRES spectrometer to achieve a certain signal-to-noise (SNR) level per pixel near 5500 Å. For a few of the faintest stars, notably some of the fainter targets in the KFOP program, the exposures were limited by a maximum exposure time, typically 20 or 30 minutes, achieving a somewhat poor SNR, but always over 25:1 per pixel. As described below, the effective noise in our search for laser emission comes from the absorption lines in the stellar spectrum, not the Poisson statistics of the photons. The exposure meter stopped the exposures when a set number of photons was received. Typically, the exposure meter was set to achieve a SNR in the reduced spectrum of 100 to 200 (i.e. 10,000 to 40,000 photons per pixel), with virtually all noise coming from Poisson fluctuations in the arrival of the photons rather than from read, dark, or any other noise source.

The vast majority of the target stars observed for exoplanet research by the CPS team were brighter than Vmag = 11, yielding exposure times of 1 to 10 minutes, depending on the star's brightness. In contrast, the target stars coming from planet-transit surveys, such as the NASA-Kepler, TrES, HAT, WASP, XO or CoRoT surveys (Bakos et al. 2012, Enoch et al. 2011) had common V magnitudes of 11 to 13.5, for which we typically achieved SNR = 100 to 25 per pixel, requiring integration times of 20 - 45 minutes. Thus, integration times varied from 1 to 45 minutes to accommodate the various brightnesses of the stars and our SNR goals that were driven by exoplanet work. The CPS team used different slit lengths as needed to accomplish sky subtraction, when needed. For stars brighter than V < 9 mag, the CPS team typically used a slit ("B5") with angular dimensions 2.5 x 0.87", preventing sky subtraction which is not necessary during the non-twilight times. Only light coming from regions surrounding the star within those angular dimensions was included in the search for laser lines to follow. For the stars fainter than V=10, the CPS team commonly employed a long "C2" slit, 14" x 0.87", to allow sky-light contributions (light pollution from city lights or scattered moon light) to be subtracted from the spectrum in the reduction process. Such background sky-light comes from light pollution from city lights, scattered moon light, and atomic and molecular emission from the molecules in the Earth's atmosphere.

In our previous laser search (Tellis & Marcy 2015), we limited ourselves to using only those spectra taken with the "C2" entrance slit (providing us with spectra from the dark sky surrounding the star), but when dark sky is not the objective we have no such limitation. In this search, we used the spectra as reduced by the CPS team rather than the raw CCD images for the initial detection of the candidate laser emission. The result is that the reduced spectra taken with the C2 slit have had the sky contribution removed but the others taken with the B5 slit have not, introducing night sky emission lines as a false positive source.

For the exposures taken with the long C2 slit, the reduced spectra only included the light coming from the seeing profile out to the 99.5 percentile level in the wings, which typically extend ~2" on both sides of the star. Thus, for both slit sizes, B5 and C2, the light included in our analysis of the reduced spectra came from within +-2 arcsec (the extent of typical far seeing wings) of the star along the length of the slit and from within +-0.44 arcsec along the width of the slit, in half-width. For all exposures, the entrance slit of the HIRES spectrometer was oriented with the long dimension vertical (perpendicular to the horizon, "parallactic"), and thus the orientation depends on the hour angle and declination at the time of observation, captured in the FITS header in the publicly available spectra from the Keck

Observatory Archive. The orientation of the slit would determine the exact rectangular angular domain from which the light was collected during the exposure, representing the observed angular region around the star.

No filters were used in the HIRES spectrometer during these exposures. The KV370 filter commonly used to remove light shortward of 370 nm was not employed. Thus, at wavelengths longward of 600 nm, our spectra suffer "contamination" from any UV light coming from 2nd order diffraction off the cross disperser grating having exactly half the wavelength of the conventionally desired 1st order light. For the stars studied here of spectral types FGKM, that contamination from UV starlight is less than a few percent simply because stars cooler than 6500K emit so little UV light. However, UV emission lines, including any of artificial origin, with wavelengths longward of 300 nm would appear in our spectra overlapping the stellar spectrum at the nominal location of (first order) light having twice the wavelength of the UV laser light. Thus our spectra longward of 600 nm are doing double duty, sensitive to laser light at the nominal first-order wavelength and at half that wavelength. Light shortward of a wavelength of 300 nm cannot penetrate the Earth's atmosphere, leaving the spectrum shortward of 600 nm free of any UV contamination.

Many of the spectra were taken with an iodine cell in place for precise radial velocity measurements. These spectra are littered at wavelengths of 4950 to 6500 Å with thousands of narrow (FWHM ~ 5.6 km s$^{-1}$) iodine absorption lines, posing challenges to our determination of the stellar continuum and to the identification of emission lines standing above the chopped spectrum.

2.3 Spectroscopic Data Reduction

We operated (initially) on the reduced spectrum to keep the volume of data from 67000 spectra manageable. Only after the first candidate laser emission lines were identified did we revert to detailed examination of the raw CCD images. For our purposes, wavelength on the raw CCD varies along a given spectral order in the horizontal "rows", whereas a vertical "column" within a spectral order is the footprint of the slit – i.e. it represents the spatial dimension. The bias level was subtracted using the median of the counts in 9 rows of each image that had no light hitting them. The locations of the spectral orders were determined by using an algorithm that detects the ridge along the middle of each order, typically ~3 raw pixels wide FWHM, and determines the displacement in the spatial direction of each order relative to its nominal location, used for later extraction of the counts at each wavelength within the order.

If the 14 arcsec long slit, either C2 or B3, was used in the observation, the sky brightness is determined at each wavelength (i.e. at each pixel along the spectral order). The medians of the outermost

~5 pixels (2 arcsec), well separated from the stellar order on either side, are determined and averaged to constitute the "sky" level at that wavelength, and subtracted (column-by-column) from the raw spectrum before rows are co-added in the reduction process. The (on-chip binned) pixels in the spatial direction span 0.38 arcsec. This sky-subtraction step effectively removes both telluric emission lines (i.e. from atoms, ions, and molecules in the Earth's atmosphere or from city lights) and continuous emission (i.e. from scattered moonlight, city lights, and nebulae) that uniformly illuminate the full 14 arcsec length of the slit. Any light (line or continuous) coming from an unresolved "dot" within 2 arcsec of the star, i.e. a machine physically located near the star, would not be removed. Thus any laser emission from within 2 arcsec of the star is retained in the final reduced spectrum.

The raw CCD images are flat-fielded by using a series of ~50 exposures of an incandescent lamp, co-added. A smoothed version of this co-added set of 50 exposures is constructed by determining the median value along each row (direction of dispersion) plus or minus 30 pixels. We divide the original co-added set of 50 flat-field exposures by the smoothed version to create a final flat-field exposure that retains the pixel-to-pixel variations caused by differences in quantum efficiencies of the pixels, but having an average value of 1.0, i.e. "normalized at unity". We divide each CCD image by this normalized flat-field, thereby removing the effect of differing quantum efficiencies of the pixels but retaining the approximate number of counts in each pixel. The counts are converted to electrons that represent photons by using the gain for the three CCDs which was set to 2.0. All figures that appear in this paper that display raw CCD images choose the bias voltage as the zero point of the colormap used.

High energy particle events ("cosmic rays") are subtracted from the raw CCD image by identifying individual pixels that are more than 5 standard deviations above their neighboring pixels. Such discontinuities cannot occur for light that passes through the optics of the spectrometer with its point spread function of approximately 4.3 pixels. This removal is quite conservative: cosmic rays that enter the CCD at glancing angles can deposit electrons in two or three (or more) adjacent pixels, looking somewhat like tracks in a spark chamber, and will often be retained and can masquerade as laser emission lines, to be discussed later.

One dimensional spectra are extracted by summing the counts along the spatial direction at each wavelength such that 99.5% of the seeing profile is included, typically 8 to 12 pixels total, symmetrically about the center of the order in the spatial direction. Thus, all light is included in the reduced spectrum within 4 to 6 pixels (each 0.38 arcsec) of the star, corresponding to 1.48 to 2.22 arcsec from the star, depending on the seeing.

Finally, the blaze function of each spectral order is removed by an algorithm that uses 50 past spectra of rapidly rotating B-type stars. Over the ~100 Å of each of the spectral orders of these stars, the intensity of light is constant within a few percent. The observed decline (of over 50%) in counts toward the ends of each order, ignoring the few Balmer absorption lines and other lines, effectively conveys the efficiency of the spectrometer (blaze function) along each spectral order. Each spectral order of the resulting reduced spectra therefore has a continuum brightness that is flat to within a few percent.

## 3. Identifying Laser Emission
### 3.1 The Properties of Detectable Laser Emission

An extraterrestrial optical laser viewed from Maunakea would be unresolved spatially, yielding an image with a shape given by the point spread function (PSF) of the Keck telescope, just as with the image of the target star in optical light. Contributions to this spatial PSF are dominated by the time-dependent refraction through the Earth's atmosphere (seeing), which typically has a FWHM of between 0.6 and 1.2 arcseconds (and occasionally outside that range). In the wavelength domain we depart from our previous program (Tellis & Marcy 2015) in which we limited the search for prospective laser candidates to those exhibiting an emission line width in wavelength closely similar to the resolving power of HIRES, i.e. the instrumental profile of the HIRES spectrometer. Typical modern continuous wave lasers are limited in their line width by coherence length, Doppler effects of the lasing atoms, and mechanical stability of the laser cavity. This yields standard linewidths smaller than 1 GHz, i.e. monochromatic with $\lambda/d\lambda > 1\times10^6$, beyond the resolution of our Keck HIRES spectra of R= $\lambda/d\lambda$ = $6\times10^4$. Pulsed dye lasers have much larger linewidths and could be resolved by the HIRES system.

In this search we demand that candidate lasers exhibit a lower bound on the linewidth of laser emission at the point spread function of the HIRES system (R=60000). However, we set no upper limit on the acceptable linewidth of features, allowing for the possibility of a series of laser emission lines that overlap when "smeared" out by the point spread function of the HIRES optical system, or any broader-band emission feature. We do not prescribe an upper bound on the linewidth, but linewidths that exceed approximately 200 pixels FWHM (~240 km s$^{-1}$) would increase the threshold required for a candidate detection, thereby making such lasers stand out less clearly than those with similar amplitude and lower linewidth. This is a consequence of the particular thresholding regime adopted, described in Section 3.2. This relaxing of the acceptable bandwidth allows us to detect a series of laser lines or an arbitrary broad and non-astrophysical emission feature, though it makes our job of detecting laser lines more difficult, due to plentiful monochromatic emission with kinematically or thermally broadened lines from naturally occurring astrophysical sources (Johansson & Letokhov 2004, 2005).

The 2-D point spread function of the HIRES system has a FWHM of 2.0 to 4.0 pixels in the spatial direction, each pixel spanning 0.38 arcsec in the spatial direction, and 4.3 pixels (FWHM) in the wavelength direction, varying by ~0.2 pixels over the echelle format due to the optics of HIRES. Candidates with too small a width in the spatial direction did not come through the Earth's atmosphere and those with too large a width are extended, resolvable by the Keck telescope and hence not an extraterrestrial laser.

There are multiple teams working on compelling pulsed laser SETI efforts, with the ability to detect individual short-duration pulses. Our method has essentially no temporal resolution and thus it detects isolated pulses only if their total energy exceeds our adopted threshold, despite our exposure time being much longer than pulse duration. For example, a nanosecond pulse containing 1000 photons at the Keck telescope could be detectable here as an emission line, despite the lack of time resolution, as CCDs integrate the total photon flux with no dead time. A pulse duty cycle of as little as $1 \times 10^{-6}$ would result in a flood of photons at Keck. Indeed, for efficient data transmission, we could expect an even higher duty cycle (see Section 6).

### 3.2 Laser Emission Search Algorithm

Each exposure taken of every target was analyzed separately for laser lines according to the same methodology, acting on the reduced, 1D spectrum. The removal of the blaze function, explained above, allowed us to develop an algorithm for detecting laser lines having an intensity greater than a defined photon threshold above the stellar continuum flux. Any emission standing above this value constitutes possible laser emission. Over the wide range of target types in our survey, including different types of stars, many or most have significant absorption features in many spectral orders that "pollute" the otherwise smooth stellar continuum.

Conceptually, our algorithm for detecting laser emission was kept simple by design, to enable follow-up validation of the laser candidates by human inspection to weed out rare artifacts. We defined the stellar continuum height in photons per pixel, $C$, to reside at the 90th percentile of the photons in a particular spectral order. This choice of 90th percentile stemmed from tests involving the flux variations caused by both stellar and iodine absorption lines, yielding a level within a few percent of the true stellar continuum despite such lines. To search for laser lines, we developed a simple preliminary sieve to quickly identify plausible candidates suitable for further analysis. This sieve was based on the notion that statistically significant laser emission must raise the flux above the continuum by more than the effective noise level and must extend over a wavelength segment at least as long as the FWHM of the instrumental profile of the spectrometer, since the emission (even if intrinsically monochromatic) will be smeared in wavelength by the optics. We defined a metric for this sieve as follows.

In each spectral order we identify all sets of at least 3 consecutive pixels (bins of 3 pixels in length) that exceed the continuum level, $C$. We chose a length of 3 pixels because the FWHM of the HIRES instrumental profile is 4.3 +- 0.2 pixels over the entire echelle spectral format. This instrumental profile implies that statistically significant emission lines must have 4.3 consecutive pixels, and certainly more than 3 pixels, with flux above the continuum. Any portion of the spectrum having only 2 consecutive pixels above the continuum level is likely to be simply a statistical fluctuation or cosmic ray hit, commonly occurring. The use of 3 pixel bins rather than 4.3 pixels allows for fluctuations at the edge of a narrow emission line that happen to drive one pixel below the continuum. For each of these sets of pixels, we record the median value, and within the entire spectral order compute the median of these medians, $m$, a metric of the typical positive deviations above the continuum for three consecutive pixels. These positive deviations are caused by both Poisson fluctuations and (more commonly) by spectral regions between absorption lines that reside above the 90th percentile continuum, $C$, defined above.

We carefully establish a threshold, $T$, for these medians of the sets of three pixels based on the median of medians. We ran thousands of analyses of our actual spectra to determine a multiplicative factor, $n$, for the median of medians, $m$, constituting a level above the continuum that these sets of 3 consecutive pixels only rarely exceeded by mere fluctuations (noise). This preliminary threshold for candidate emission lines is given by $T = C + n*m$ where $n$ is some positive value chosen to allow an acceptable number of false positives, described below. The value of $n$ is analogous to a conventional threshold based on the number of standard deviations above the background continuum required to trade false alarm probability with detection completeness.

A few percent of our spectra, or spectral orders, have less than ~100 photons per pixel, either because the star is faint, the exposure was short, or the red color of the star left the far blue portion of the spectrum with little flux. In these rare cases, the Poisson variations become comparable to the flux variations from the absorption lines, roughly 10 to 30%. However, the number of photons cannot be negative. This Poisson positivity bias leads to fluctuations commonly above the nominal value of $T$, leading to an increasing number of false positives with decreasing photons per pixel. For these rare, low photon level cases we assign an upper bound on the threshold of $T = C + n * m * 3/2$, which artificially raises the threshold relative to the continuum, making the criterion for identifying lasers more conservative. The result is to limit the number of false positives. This case of under 100 photons per pixel affects only a few percent of our spectra, and occurs primarily at wavelengths less than 4000 Å where late-type stars have little flux. We note that these cases of artificially raised thresholds actually allow laser emission to be more easily detected due to the low competition from the star light.

We record all sets of at least 3 consecutive pixels that are above the threshold, $T$, deeming them preliminary candidates We took care to rule out candidates at the wavelengths of common stellar emission lines such as the Balmer series occurring in magnetically active M and K dwarfs (as well as many uncommon lines), and night sky emission wavelengths from atomic and molecular species in the Earth's atmosphere and from city lights. Laser emission at those wavelengths will be missed. The preliminary candidate extraterrestrial laser lines were identified with no consideration of the wavelength profile shape of the emission. Near the threshold, the effective noise in the underlying spectrum (much of which is due to myriad atomic absorption in the star's atmosphere or from molecular iodine lines in our local cell) is too poorly known to demand precise profile shapes, other than that at least 3 consecutive pixels be above our threshold. As described previously, we imposed no upper bound on the wavelength width of a candidate laser feature. A series of laser emission lines within a wavelength range could easily overlap due to the smearing by the PSF of the Keck HIRES optical system, appearing as one wider emission feature. For similar peak flux, a given effective wideband signal would be more easily detected, as still only 3 consecutive pixels must exceed our threshold.

3.3 False Positives: Night Sky Lines, Cosmic Rays, and Instrument Artifacts

There are a number of features commonly present on the CCD image that will sometimes be mistaken for laser emission by our algorithm. Note first that we ignore entirely the shape of the laser emission with wavelength, as this would necessitate subtracting out with high fidelity the contribution of the underlying spectrum with its stellar and perhaps iodine absorption lines. Each of the following features was easily mitigated, but individual corrections were made for each.

First, many OH- and $O_2$ night sky emission lines are present in the spectra, especially redward of 5500 Å. In spectra that used the long 14 arcsecond slit, sky subtraction was performed, preventing these sky lines from being retained in the reduced spectrum. However, for the other spectra, it was necessary to ignore any laser line candidate that fell within a few pixels of the peak of a well-known night sky emission line. The wavelengths used for this mitigation were identified with the Keck 1 telescope, providing a good match to the emissions present in the sky over Maunakea (Osterbrock et al 1996). The catalog used to remove these emission lines contains lines of varying strengths, and only those that commonly exhibit flux sufficient to be detected as a laser candidate in the reduced spectrum were removed.

Second, the CCD images are littered with the "cosmic rays" and "worms", these being the names for muon tracks through the CCD and the paths of multiply-scattered low-energy electrons liberated by gamma rays, respectively. Under some specific circumstances, the conservative cosmic ray removal component of the reduction pipeline does not remove these, especially the worms. As we do not on the first pass consider the shape of the prospective laser signal, these are reported as candidate laser lines when they illuminate at least three consecutive pixels, i.e. when the muon track moves through the CCD at a grazing angle, or the worm is sufficiently long and bright. It is trivial for the eye or a simple algorithm to tell that these signals cannot possibly correspond to light coming from an astrophysical source, as they lack the telltale smooth drop-off associated with the instrumental profile of HIRES. In order to identify these signals after the first pass, we modeled the spatial profile of the stellar ridge using the sum of the counts in 20 columns along the stellar spectrum located before and after the 11 columns surrounding the candidate signal. We then computed a simple reduced chi-square statistic to the signal columns, setting a very lenient threshold of $\chi^2_{red} < 25$ in order to be retained. If this chi-square statistic was above 25 in one or more columns, that column was deemed polluted by a cosmic ray due to its poor fit to the spatial profile, and the candidate was rejected. Even with this lenient threshold, the process allowed us to significantly cut down the amount that would require later examination by eye.

Special care had to be taken in removing these classes of false positives, lest we somehow introduce filtering regimes that would throw out the signals that we are looking for. Additionally, every wavelength regime that is ignored for some reason presents another small band in which we are blind to laser emission. For example, we also ignore the wavelengths at which a bright diagonal stripe known colloquially as the "meteor" runs through the center of the middle CCD caused by internal reflections within the HIRES spectrometer. We had to ensure that in dealing with these false positives, we did so in a way that preserved as much of the search space as possible. These ignored wavelength domains in total account for less than 3% of the total search space, though it is of some importance that the bands ignored surround common emission lines. It could be argued that an ETI hoping to communicate with Earth would be aware of their host star's emission bands and would purposely communicate in an unexpected wavelength regime. A yet more careful analysis could determine if the widths of a target's emission lines are sufficiently tight to be consistent with a laser line, rather than a thermally broadened stellar emission.

3.4 Setting thresholds and assessing sensitivity

Two factors affect the detectable laser flux, namely the continuum flux of the stellar spectrum and its fluctuations, against which the laser line must compete. We adopt the continuum level as a specified percentile of photons per pixel, $P_{cont}$, among all pixels within each spectral order. We compute a fluctuation level based on the median of the absolute value of the deviations of the numbers of photons within bins of three consecutive pixels. The continuum level is trivial to determine in the absence of absorption lines, stellar and iodine. The machine determination of the continuum when it is chopped up by such absorption lines is less straightforward, though we ended up using a very simple approximation. We found that the $90^{th}$ percentile of the photon level per pixel provided an excellent determination of the continuum as assessed by eye for all spectral orders and all spectral types. This $90^{th}$ percentile minimizes the suppression of continuum level caused by the scattered absorption lines but leaves enough pixels in the remaining upper $10^{th}$ percentile to firmly set the continuum against the Poisson noise. Note that the exact adopted level of the continuum level is not important, as long as it is done consistently. The subsequent injection and recovery of synthetic laser lines will set the actual detection thresholds and false alarm probabilities.

Adopting the $90^{th}$ percentile of the photons per pixel within an order as the continuum level does affect the appropriate number of median absolute deviations, *Nmad*, to adopt in our threshold determination. We tested many values for *Nmad* with attention to its effect on the efficiency with which of the algorithm detected simulated laser lines. Because of the variability in the absorption characteristics of the stars observed, we found that setting the threshold at levels below *Nmad* of 6 above the $90^{th}$ percentile continuum resulted in several false positives per observation, unacceptably high, especially given 67,000 spectra to analyze. The code frequently identified natural fluctuations in the stellar spectrum itself as potential laser emission, which would have produced tens of thousands of false alarms. These false positives were clustered in the orders where the continuum was underestimated – chiefly low-wavelength orders blueward of 430 nm containing low flux and numerous stellar lines, and orders containing the dense iodine absorption spectrum. We found that setting the threshold at *Nmad* = 6 kept the occurrence of the false positives to about one in every five hundred observations, or 127 occurrences in total. Such a low occurrence rate was desired, as ultimately all laser candidates (after judicious algorithmic sieving) were assessed individually by eye to determine the plausibility of being laser emission.

In order to determine the tradeoffs of the continuum and *Nmad* thresholds, we performed a test of our algorithm by injecting simulated laser emission into randomly selected spectra in our sample and assessing how many of the injected laser lines were recovered by the code successfully. We simulated the prospective laser emission as having an instrumental profile approximated by a Gaussian with a full-width half-max of 4.2 pixels. Ten of these simulated lasers were inserted into each spectral order at random sub-pixel locations and were tested for recovery, keeping track of those that overlapped or fell on known night sky or astrophysical emission lines. These mock laser lines were inserted at 40 different flux levels set by the thresholds computed for each spectral order. This process was repeated 40 times per observation to attain a robust average, and was performed on 3 observations with the iodine cell in and 3 without for each spectral type, F, G, K, and M main sequence stars, as well as the fainter (lower SNR) spectra of the *Kepler* stars. In total, we simulated 480,000 laser lines in each spectral order for each class of star, for instance K-stars without the iodine cell in. These 480,000 simulated lasers were spread over three representative members of each spectral class and realized with 40 brightness levels, to determine the efficiency of detection and false alarm rate.

Figure 3 shows the peak brightness in continuum units of the simulated laser lines required at each spectral order for 80% of them to be recovered. Note that this peak brightness corresponds to the central pixel of the Gaussian model, were it centered on one pixel. Thus the actual detection threshold is lower than this peak value such that the sub-pixel sampling of the simulated laser line results in three pixels sitting above the threshold. The threshold brightness in continuum units for 80% recovery was computed by a linear interpolation of the results of the recovery simulations described above. Figure 3 shows the degree to which recovery efficiency of mock laser lines varied from order to order and the sensitivity to laser lines lost by the addition of the iodine cell, in the wavelength band from 4950-5600 A. For the high wavelength orders (above 600 nm) that contain few deep absorption lines, faint laser emission relative to the continuum brightness was reliably detected. However, in the spectra in which the iodine cell was used, the 80% recovery threshold was only met by lasers having a peak brightness of approximately 50% of the continuum height and higher in some cases. In the blue portion of the spectrum, below 430 nm, many observations had fewer than 100 photons per pixel, for which we adopted our secondary thresholding regime described previously, in which we demanded the laser emission peaks be in excess of the entire continuum height to be deemed detected. This high threshold implies that the algorithm is effectively poor in these cases. But this concern is only relevant where there are only a few photons detected, i.e. the blue part of the spectra of faint stars, such as M dwarfs or *Kepler* stars. Because the continuum flux is so weak in those cases the sensitivity to laser emission in terms of energy flux units is still good because the competition from the stellar spectrum is modest.

We focused on simulating lasers containing sufficiently low wavelength width so as to be unresolved by Keck. We also simulated lasers extending over a longer wavelength band, though with not the same level of granularity. Resolved lasers have the advantage of multiple three-pixel segments in which to exceed our threshold. As only one such set of three consecutive pixels is required for the algorithm to flag a candidate laser, these signals have a higher probability of detection. However, in the case that the integrated brightness of the signal is merely spread out over more pixels, the unresolved laser has higher probability of detection, as a wideband signal of the same integrated brightness will have lower contribution in each pixel and therefore may not exceed our threshold in three consecutive pixels. In borderline cases, the lower flux per pixel outweighs the greater number of 3-pixel bins that the algorithm can use to flag the signal, resulting in lower probability of detection. Features wider than 300 pixels or more will be successively suppressed by the deblazing and flattening of the continuum in the data reduction process. Further, the spectral orders are 4000 pixels long. Therefore, emission features broader than 300 pixels will suffer from diminishing detectability up to 4000 pixels at which point they becoming completely undetectable.

See Sections 5 and 6 for a discussion of what this dependence of sensitivity on wavelength means in terms of flux density at Earth and emitted power from a benchmark ET laser.

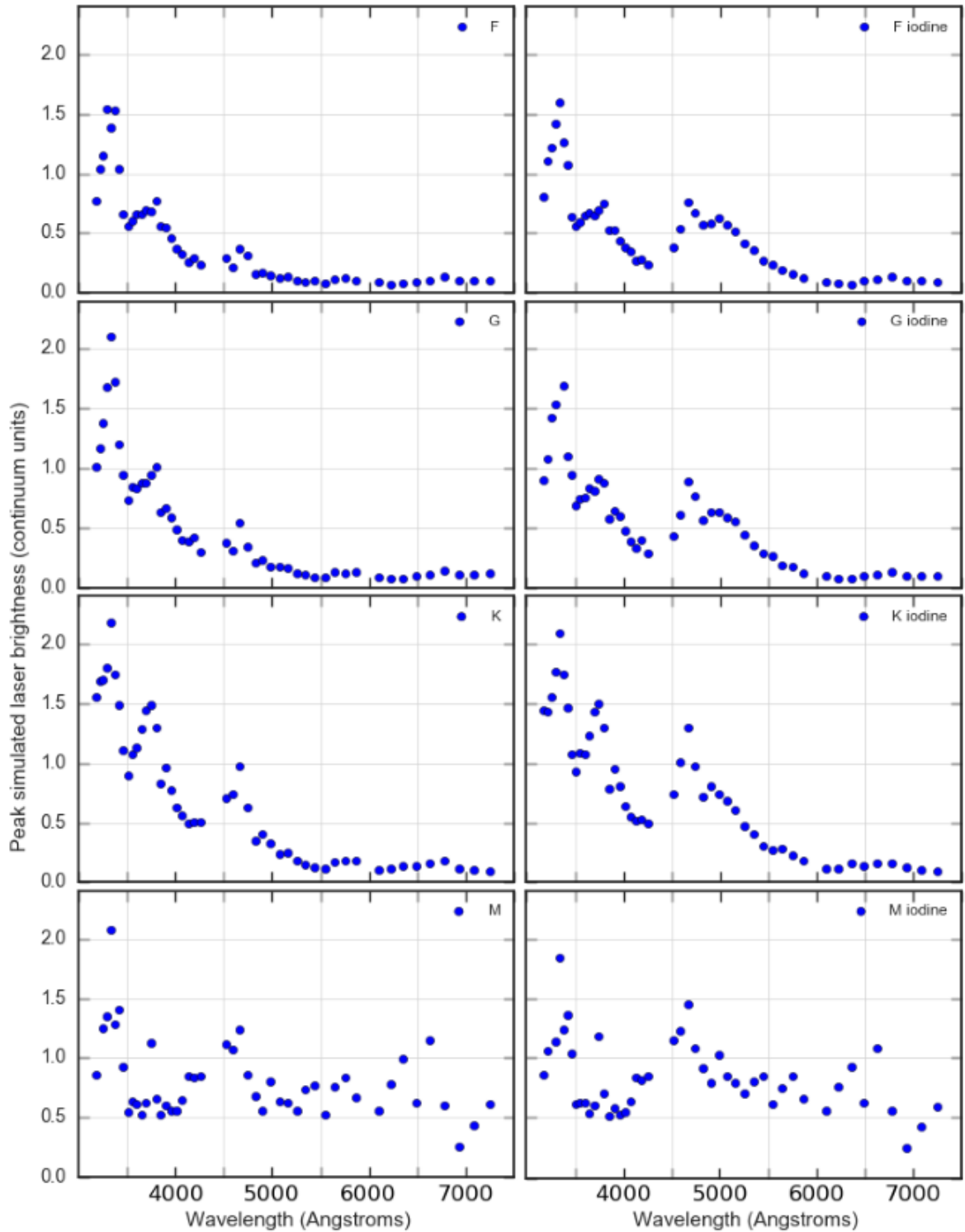

Fig 3. The required peak brightness of simulated laser lines, relative to the stellar continuum in each spectral order, for 80% to be detected, as determined by direct injection of mock laser lines and their recovery (or not) by the detection pipeline. Each panel represents a different stellar type observed without the "pollution" of iodine lines (left column) and with iodine lines (right column), which compromise the sensitivity to laser lines. Low luminosity M-dwarfs with their many atomic and molecular absorption features suffer from high detection thresholds in continuum units, yet due to their low overall flux allow the detection of faint lasers.

4. Results

Using the algorithm described in Section 3.2 we analyzed 67708 spectroscopic observations of 5600 targets for signals that could be consistent with an extraterrestrial laser. We recorded any occurrence of three consecutive pixels that exceeded the spectral order's 90th percentile count value by 6 * *Nmad*. The choice of these parameters and the tradeoffs associated with their choice are explained in Section 3.3. Executing the algorithm yielded 5023 laser-line candidates. These candidates may be divided into three main categories, namely candidates resulting from 1) small defects in the spectrometer or the spectrum reduction pipeline (poor treatment of the 2D CCD image), 2) cosmic ray, gamma ray, or muon hits on the CCD, and 3) real spectral peaks arising from emission lines of the target star or from Earth's atmosphere. Examination of the original raw 2D CCD image, as well as the reduced 1D spectrum, allowed discrimination of false positives from viable candidate laser lines.

Within category 1, false alarms can occur when the continuum of the spectrum is not flat within a percent as it should be over the small, 10 nm wavelength segment of spectrum. In such cases, the stellar spectrum itself can sometimes artificially rise above (or an emission feature fail to rise above) the detection threshold due to the non-flat continuum level. Erroneous departures from flatness can occur when the CCD image reduction code fails to locate accurately the central ridge along a spectral order. Departures can also occur when the continuum efficiency blaze function (by which the raw spectrum is divided) is inadequately determined, especially at the ends of an order where the spectrometer efficiency is falling rapidly and nonlinearly. Such non-flatness in the continuum is reliably identified by its repeated occurrence at the same locations in certain stellar orders, as well as by simply inspecting a plot of the reduced spectrum to reveal the wavy (rather than flat) continuum level. Repeatable internal reflections, i.e. ghosts, from idiosyncratic optical effects in the HIRES spectrometer also occasionally are identified as emission lines, easily rejected as viable emission-line candidates by their repeatability. Identifying and removing these category 1 false alarms brought the total number of candidates down to 2356.

Within category 2, laser-line false positives caused by cosmic rays, radioactivity from the observatory, or gamma rays were identified by our laser-search pipeline in the reduced 1D spectra, but subsequently examined by eye in the original 2D CCD image. Cosmic ray, gamma ray, and muon hits make a distinctive pattern in the raw CCD image and hence were easily deemed inconsistent with a candidate laser line. Specifically, any isolated single or double bright pixels standing alone without similarly bright neighboring pixels (befitting the optical PSF of the spectrometer) was a sure sign that the counts (electrons) in the pixels were not due to photons that passed through the optics of the telescope. Some of these high energy particles and gamma rays leave a "track" in the raw CCD image that is many pixels long, but only one or two pixels wide, and often trailing along a diagonal path, betraying its arbitrary direction through the CCD silicon substrate. In total, out of the remaining 2356 candidates, 1575 candidates were due to cosmic ray hits, gamma rays, or muon tracks.

Within category 3 were false positives consisting of narrow and rare (thus not rejected by the code) emission lines coming from the star itself. Both common and rare atomic transitions were detected in emission, especially in the magnetically active stars, emission-line stars, and variable stars. For each candidate laser emission line that could not be ruled out by explanations within categories 1 and 2, we considered the nature of the star itself, and any background nebulosity, as a possible source of the emission. The nature of the star was gleaned from any past papers written about it (found using SIMBAD) with special attention to past spectroscopic observations that revealed emission lines. Any candidate laser line having a wavelength near a previously reported or commonly occurring stellar emission line was suspect as simply arising from the star.

Any stellar emission lines usually arise from the chromospheres or coronae surrounding the photosphere, producing a common set of emission lines from the usual atomic species inhabiting the hot region. In such cases, we searched for the other commonly associated emission lines to confirm (or reject) the hypothesis that the emission line identified by our algorithm was simply one of many lines associated with the star's hot gas. Of course, one might wonder if an advanced civilization could fool us by purposely beaming laser emission lines having wavelengths consistent with those that arise naturally in chromospheres of stars. It would be difficult for us to distinguish such purposeful, nefarious camouflage from naturally occurring chromospheric lines. Indeed, all stars exhibiting emission lines in their spectra could be interpreted as laser emission from shy civilizations attempting to hide among the chromospheric weeds. If so, we throw up our hands and interpret a set of emission lines known to come commonly from gas of 5000-1000K as simply chromospheric, and not machine-born. However, we did not measure the relative brightness of these consistent emission lines. If a candidate laser line consistent with the wavelength of a known stellar emission line were brighter than expected based on emission lines from other species in the spectrum we would not know of this inconsistency, resulting in our unfortunate rejection of it as laser emission.

The figures following in this section show representative candidate laser lines identified by the search pipeline. Each candidate figure is composed of three panels. The top panel shows in blue the entire reduced spectral order in which the potential candidate laser line was found. The thresholds assigned by the code are displayed as horizontal lines overlaid on top of the spectrum. Black represents the 90th percentile value of photons per pixel (the "continuum") and red represents the threshold $T$ described in Section 3.2 above which three consecutive pixels must reside to be deemed a candidate laser line. The middle panel shows a detailed view of the spectrum in a wavelength segment centered on the candidate laser line. The black and red horizontal lines are the same as in the upper panel. The lowest panel in each figure shows the raw CCD image at the pixel location of the candidate, centered on the same pixel as the middle panel. Photon count values in the raw CCD pixels are denoted by color according to a color map displayed on the right of the figure. The maximum and minimum values correspond to the maximum and minimum CCD pixel values in the region displayed. The bottom two panels have nearly the same pixel scale in the wavelength direction, so one column in the raw image (bottom panel) corresponds to one point (from the spectrum reduction described in Section 2.2) in the middle panel. These candidate laser lines show the types of signals that were typically identified by the algorithm, as well as a few of the more interesting cases.

## 4.1 Candidates Lacking an Instrumental Profile

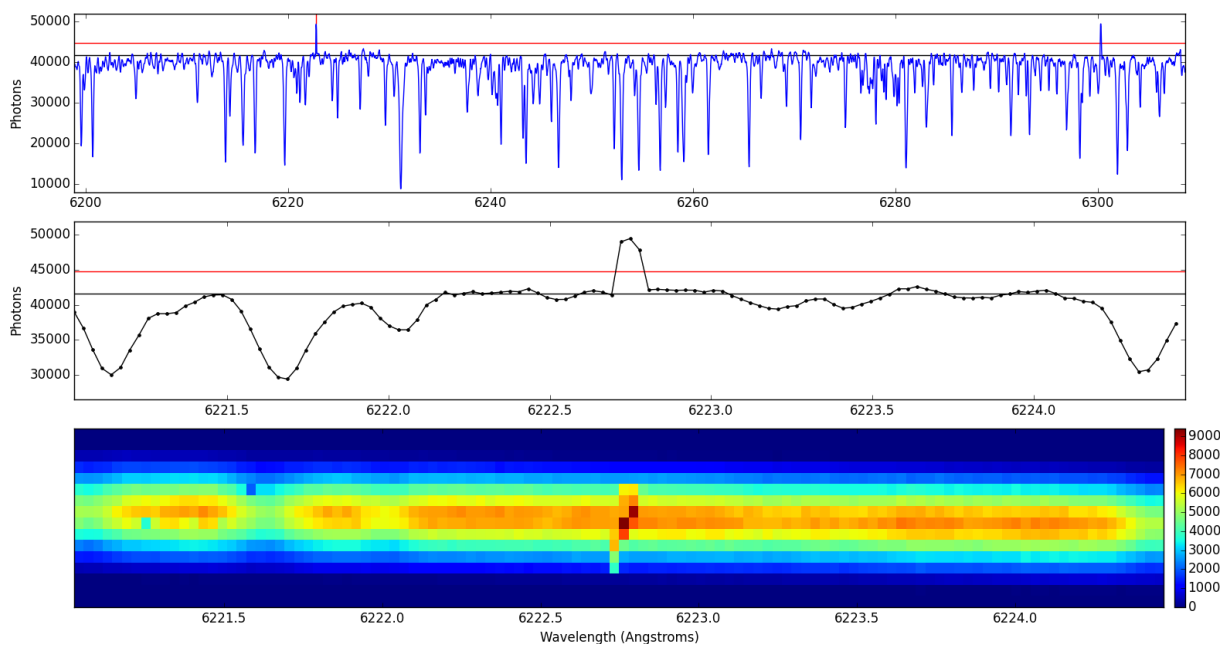

Fig 4. Candidate laser line marked by our search algorithm in HD 5319. The top and middle panels show the entire reduced spectral order in which the candidate was found and a detailed view of the area around the candidate, respectively. The black horizontal lines show the 90th percentile photon count value and the red show the threshold $T$ set by the algorithm. The bottom panel shows the raw CCD image at the same horizontal pixel scale as the middle panel, with corresponding columns.

Photons per pixel in the CCD image are denoted by the colorbar on the right. The narrow, diagonal track in the bottom panel indicates that the candidate laser line was caused by a high energy particle or photon piercing the CCD substrate.

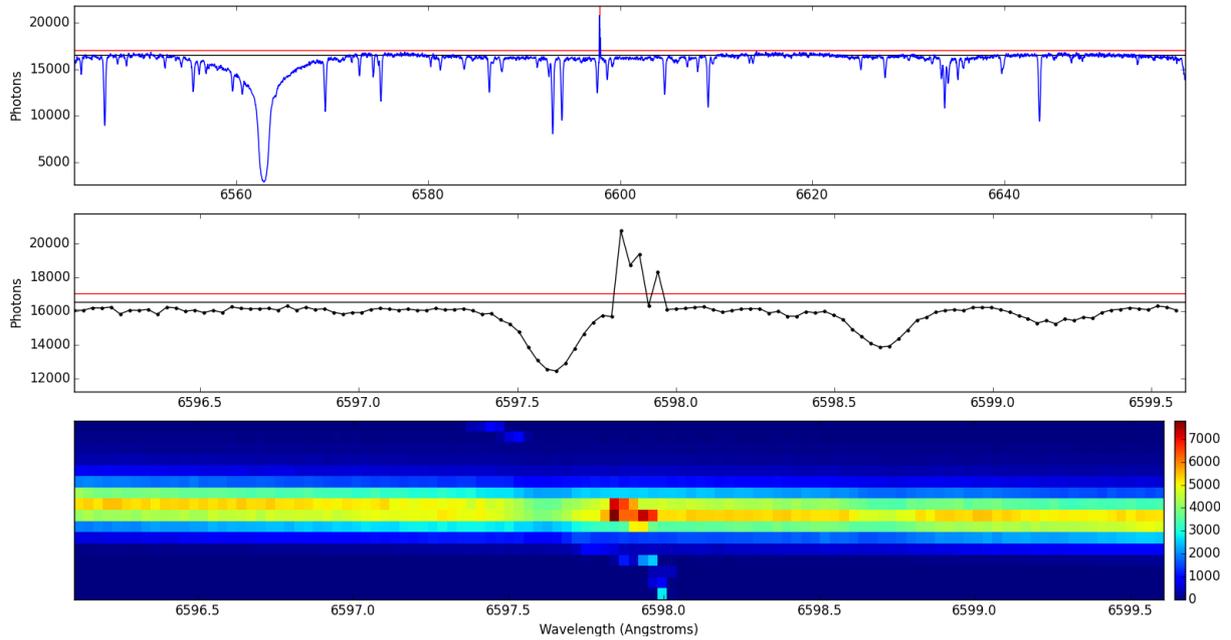

Fig 5. Candidate marked by algorithm in HD101904. See Fig 4 and the text for an explanation of the panels. Here, as in Fig 3, the diagonal track in the raw CCD image (bottom) strongly suggests a high energy particle caused the apparent "emission line" in the reduced spectrum (top and middle).

Figure 4 shows a candidate laser line identified by our search pipeline in HD 5319. While promising as a candidate in the reduced spectrum, the counts in the raw CCD image are deposited in such a way as to be inconsistent with a laser - namely there is no visible "smearing" of the signal caused unavoidably by the optics of the telescope and spectrometer. This inadequacy in line width is clear to the naked eye, though the reduced spectrum contains sufficient counts in the three consecutive pixels as demanded of candidate laser lines by the algorithm. These counts likely come from a cosmic ray muon (or perhaps other particle) impacting the CCD at an angle.

Figure 5 shows a candidate emission line in HD101904. The peak of the emission line is jagged, unlike the smooth curvature of the instrumental profile of the spectrometer. Fitting a stellar image-calibrated spatial PSF, modeled as a Gaussian, and performing column fits to the columns with elevated counts, shows a poor fit. The naked eye senses the same, and the extra counts deposited from top-left to bottom right in a semi-linear fashion are an indicator of electrons in pixels arising from a cosmic ray or gamma ray.

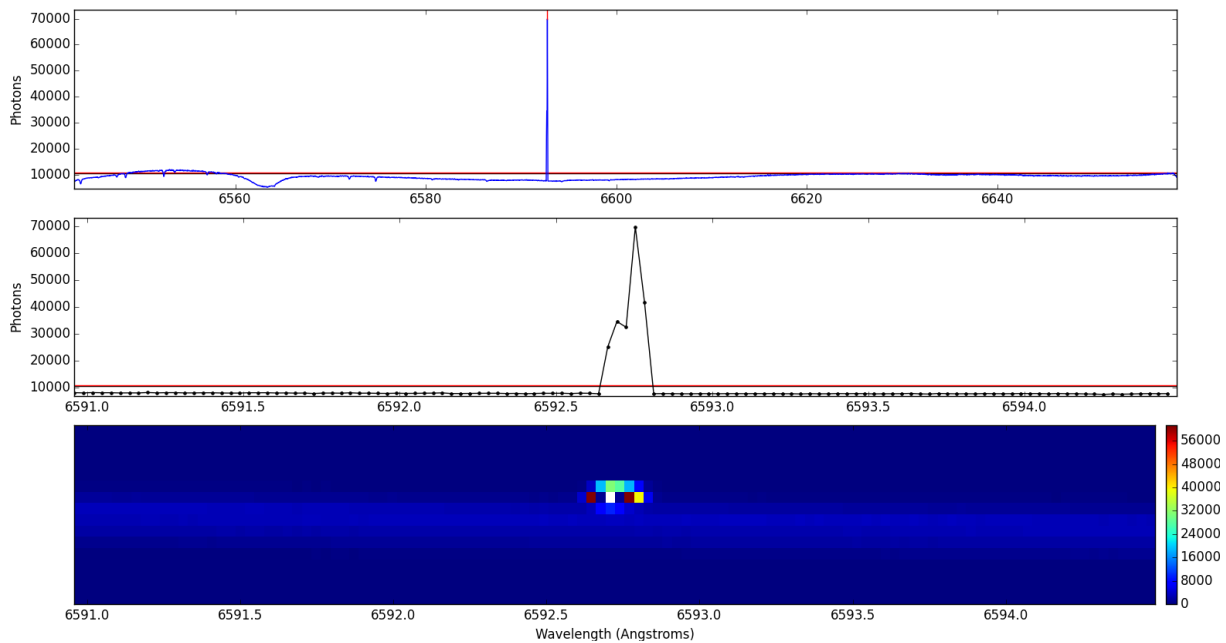

Fig 6. Candidate marked by algorithm in KIC 8462852. The white pixel in the middle is a NaN value, indicating the pixel contained more than the 65536 maximum allowed value by HIRES readout electronics. See Fig 4 and the text for an explanation of the panels. The distribution of photons in both the spatial (up-down) and wavelength directions are narrower than the point spread function in those two directions, indicating that a high energy particle directly deposited large amounts of energy on the heavily saturated single pixel (with the NaN value) which spilled into neighboring pixels. Note the absence of wings in the spatial direction that should extend 3-5 pixels from the center of the "line" if it were caused by light actually passing through the HIRES spectrometer (See Fig 7).

### 4.1.1 KIC 8462852

Figure 6 shows a candidate laser emission line found in a spectrum of KIC 8462852. Of several observations of this star, this was the only candidate returned by the laser search pipeline and so is examined here with extra care. An examination of the distribution of counts in the raw CCD image reveals pixel saturation and its effect on the electrons spilling into neighboring pixels and on the CCD readout process. The white pixel shown in the raw CCD image in Figure 6 is recorded as "NaN" (not a number), which occurs when the voltage in the pixel exceeds the capacity, 65536 counts, of the analog-to-digital converter during readout of the CCD. It is likely that the large number of electrons deposited in that pixel bled into neighboring pixels. This form of saturation is inconsistent with the saturation effects from bright monochromatic light that passes through the spectrometer, as such light would be smeared by the spectrometer optics with the usual instrumental profile, which is not present here. It is not clear exactly how the CCD readout process would produce such a pattern in the case of saturation from a cosmic ray hit. However, for comparison, Figure 7 shows the spatial extent of saturation as a result of light passing through the CCD, in this case an exposure of a thorium-argon lamp used for calibration, and compares it to that seen in Figure 6. The pixel values are displayed on a log scale to emphasize the difference between the two sources of saturation. Some striping or "bleeding" (photons spreading into neighboring pixels along a row or column) may be seen in the ThAr exposure, as well as a clear instrumental profile. Neither is present in the saturation seen in Figure 6, ruling out this high number of counts in one pixel as being due to light that entered the aperture of the telescope and spectrometer.

Finally, Figure 8 shows the overlaid profiles of the underlying stellar spectrum and the candidate seen in Figure 6. In order to compute the profiles, the spectrum and the feature were summed along the wavelength direction. The stellar spectrum profile was obtained from summing ten consecutive columns to the left of the peak of the feature, whereas the feature profile was generated by subtracting the median of each row (i.e. subtracting the star) and summing ten columns centered on the feature. The "laser" feature profile is noticeably narrower than that of the profile of the stellar spectrum, and hence narrower than the instrumental and seeing profile. Furthermore, the NaN value in the middle row of the feature was treated as zero in the sum. If this NaN value is due to a readout error and some nonzero number of counts are present in this pixel, this pixel actually contained many more electrons and hence the profile is even narrower that it seems in Figure 8, i.e. even less consistent with the true spatial seeing profile.

Figure 9 shows another saturation event, this one in CK01474 (KOI 1474). The two long tracks of counts suggest this particular saturation may be due to a pair of charged particles produced from one impacting cosmic ray muon (or other particle) that hit the CCD at a lower angle than for the candidate in Figure 5. The eye can tell right away that this is not light that passed through the spectrometer, however in three consecutive columns the cosmic ray removal pipeline failed to detect and remove this. Again, the form of saturation here, and the tracks, are not consistent with light passing though the spectrometer.

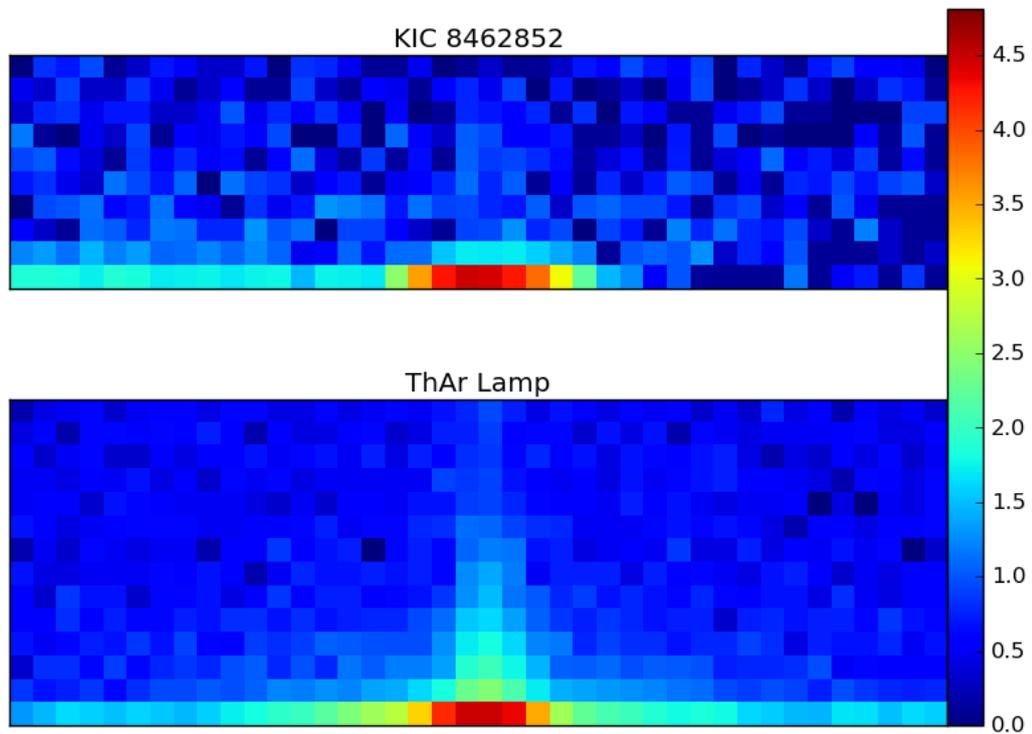

Fig 7. Raw CCD images showing the effect of saturation. At top is the candidate laser line from Tabby's star, showing the lack of upward spatial extent. This was shown in Fig 6, but here the color of a pixel scales with $\log_{10}$(counts/pixel). At bottom is a saturated thorium line from a spectrum of the thorium-argon calibration lamp. The counts in the thorium line were scaled (by 10%) so that both the top and bottom plots have the same total counts. Both plots are on the same log scale, with the color code at right. The thorium lines exhibit an extension of the counts in pixels above and below the center, caused by normal scattering of photons off the cross-disperser diffractional grating. The emission feature in Tabby's star should exhibit the same extension above and below due to the same scattering of photons. But such extension is not there, indicating that the feature in Tabby's star was not caused by photons passing through the spectrometer. Thus it seems unlikely that this event in the spectrum of Tabby's star is due to laser emission.

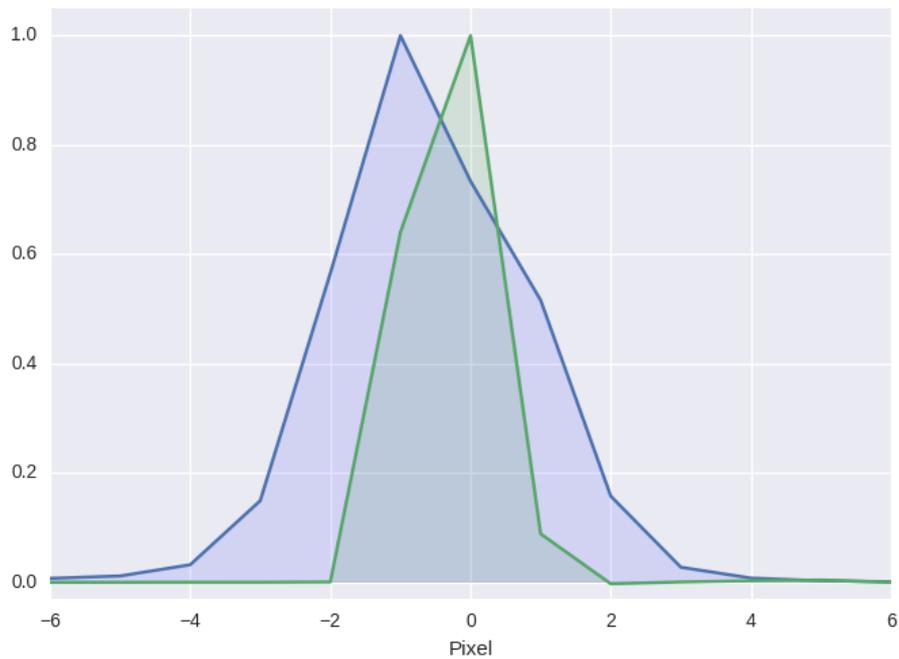

Fig 8. Comparison of the profiles in the spatial direction of the stellar spectrum (blue line) and the laser candidate feature found in Tabby's star (green line). The profile of the stellar spectrum reveals the instantaneous refraction by the Earth's atmosphere ("seeing") of a point source (the star). The laser candidate feature exhibits a profile that is demonstrably narrower than that of the star, strongly suggesting that the candidate feature does not originate from light entering the Earth's atmosphere. The stellar spectrum profile was obtained from summing ten consecutive columns to the left of the peak of the feature, whereas the feature profile was generated by subtracting the median of each row (i.e. subtracting the star) and summing ten columns centered on the feature. The peak pixels of both profiles are normalized to unity. The NaN pixel in the candidate feature was assigned a value of 0 in the sum, thus the pixel in the green profile at position "0" likely has more counts than is plotted here. If more photons were associated with that saturated pixel, the laser candidate feature would simply be even narrower than that due to atmospheric seeing, thus supporting the conclusion that it didn't originate from light that passed through the Earth's atmosphere.

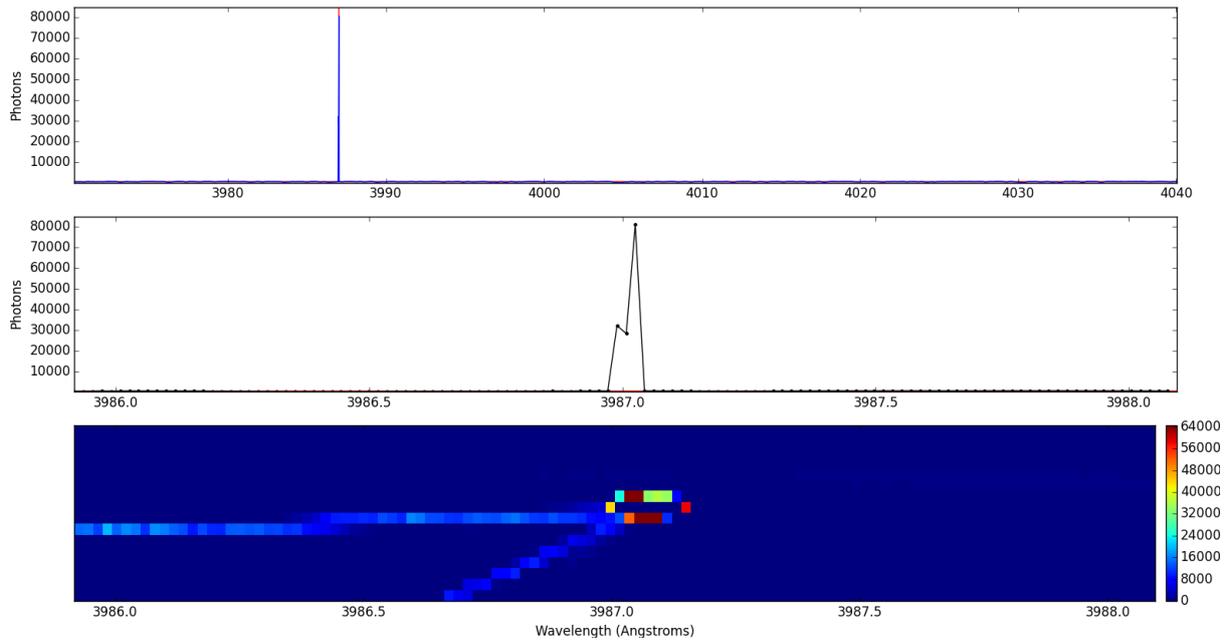

Fig 9. The Kepler candidate exoplanet-host star, CK01474 showing apparent narrow emission (top and middle panels). The bottom panel shows two tracks in the original CCD image, indicating the cause as one or more cosmic ray events saturating the central pixels. See Fig 4 and the text for an explanation of the panels.

4.2 Candidates Resulting from Monochromatic Emission

The next candidates resulted from light entering and being properly dispersed by the spectrometer. Figure 10 shows atomic emission in magnetic, spotted, flare star YZ Canis Minoris. This is a common line of Si I at wavelength 3905.5 Å. It showcases the difficulties in finding lasers amid a sea of emission lines from the star itself. Many such lines, especially in stars with strong, widespread magnetic fields, were found by the pipeline but ruled out as convincing lasers due to their match with transitions in common atoms found in these stars, or the presence in the same spectrum of other related atomic transitions. Frequently these are difficult to rule out as laser lines by themselves, due to minimal rotational and thermal broadening of the emission line.

Figures 11a and 11b show remarkable candidate laser emission found in HAT-P-39. This interesting case at first appears to show light that could be consistent with a series of laser emission lines in a picket fence of nearby frequencies. However, on further examination, the light illuminates the entire slit in the spatial direction, seen faintly in the middle panel of Figure 11b, and therefore must be due to broadly distributed, scattered light in the Earth's atmosphere. The source of this light is unknown, though the telescope on Maunakea was pointing just south of the town of Hilo at the time of exposure (Azimuth angle of 115.37 degrees relative to the Keck Observatory, and low HA). This picket-fence of emission lines was seen in exposures taken before and after, with different azimuth angles, providing further evidence that the source is terrestrial, but of unknown origin. The intensity of the lines is confined within an envelope, suggesting an origin from an optical cavity. One explanation is that this is scattered light from the adaptive optics guide laser on a neighboring telescope. If so, it would have to be from a harmonic, as the vast majority of AO systems employ sodium guide lasers. Sky subtraction should have eliminated such Earth-borne lasers when the long slit was used.

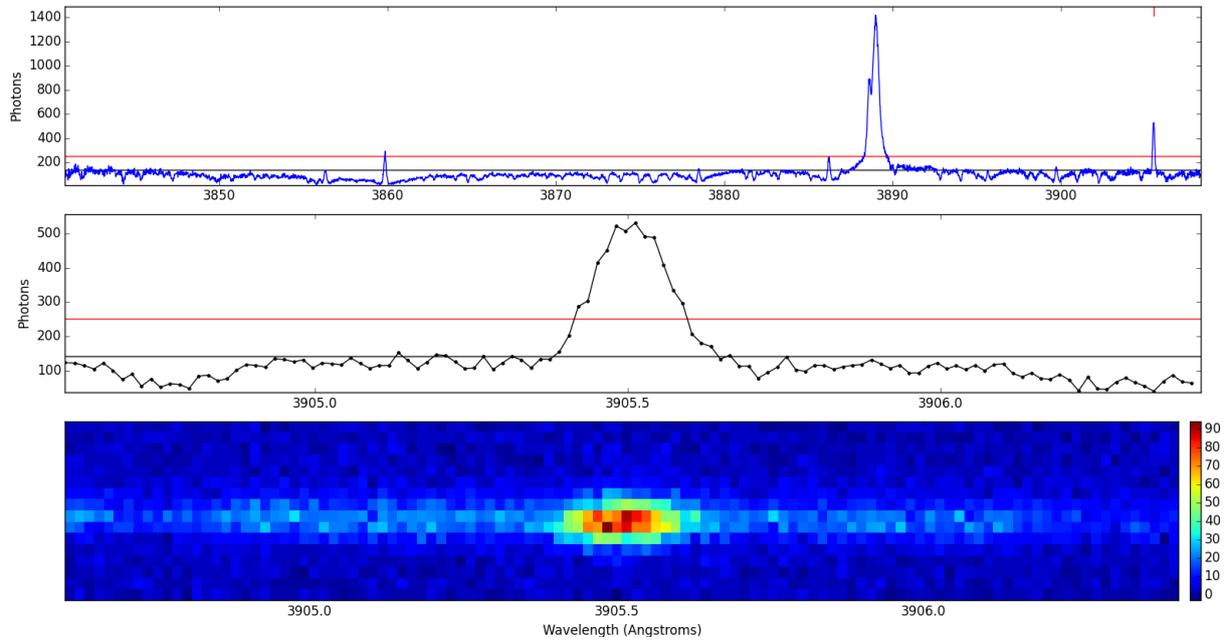

Fig 10. Atomic emission from the chromosphere in the flare star YZ Canis Minoris. See Fig 4 and the text for an explanation of the panels.

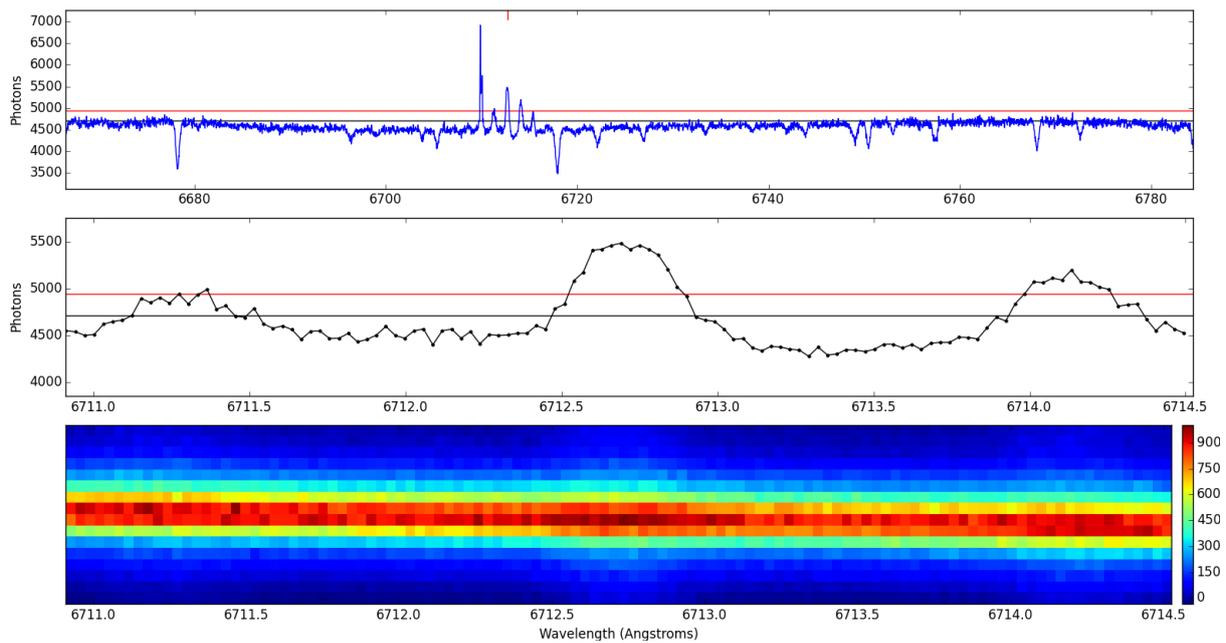

Fig 11a: Candidate found in HAT-P-39. See Fig 4 and the text for an explanation of the panels. Five emission lines having different intensities but equal wavelength spacing are apparent to the eye in the reduced spectrum (top panel). A zoom-in (2nd panel) shows three of those emission lines. The emission fills the spatial direction, albeit faintly (3rd panel). This series of emission lines is apparently unresolved emission from a laser or oscillator on Earth, the light subtending a large enough solid angle to fill the spectrometer slit.

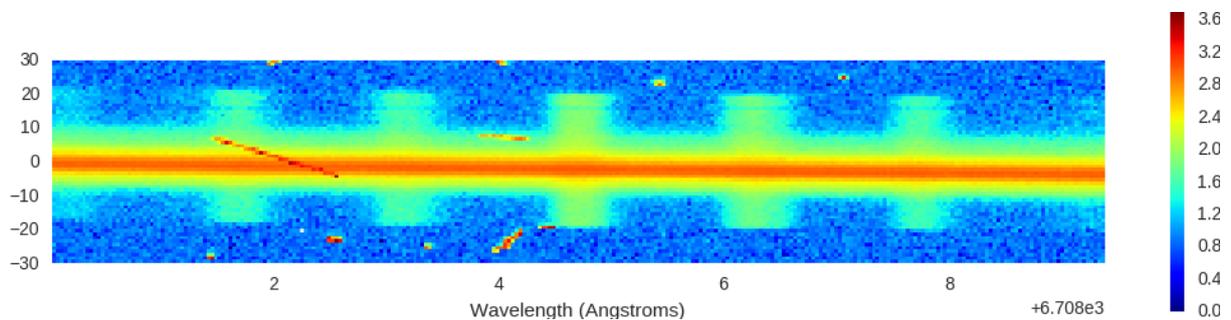

Fig 11b: A further zoomed-in view of the same feature in Figure 11a, plotted on a log scale to enhance the contrast and exhibit the line's spatial extent spanning the entire entrance slit. The diagonal line on the left is yet another glancing particle track.

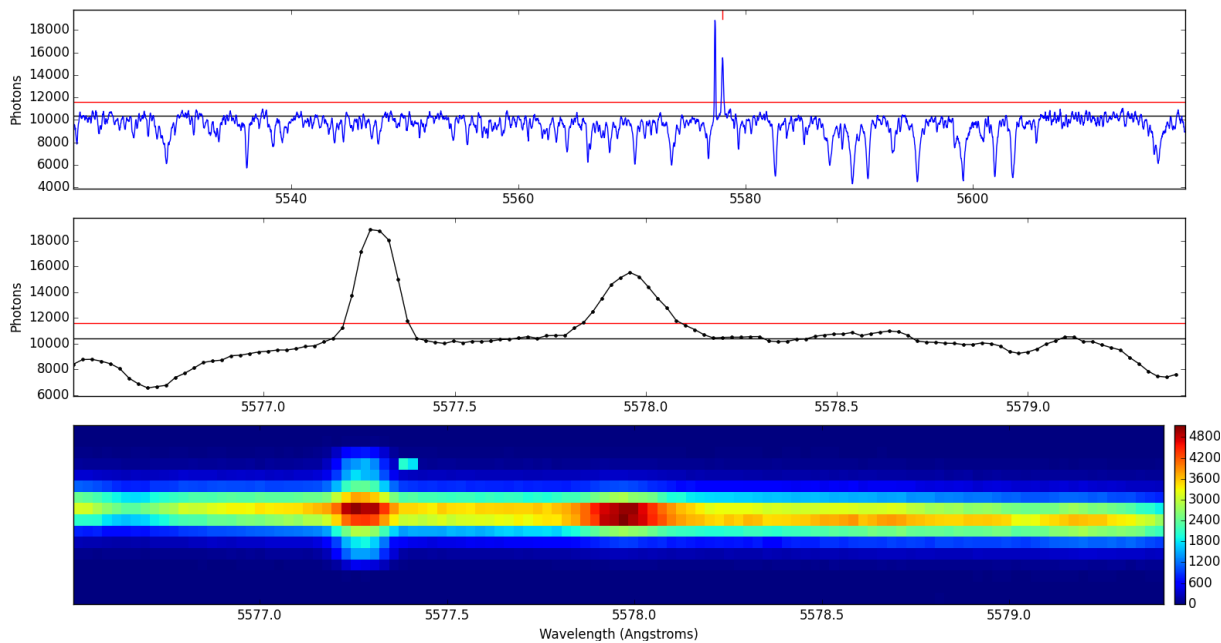

Fig 12. Extraterrestrial laser candidate found in an observation of the T Tauri star, TW Hydrae. See Fig 4 and the text for an explanation of the panels. The two peaks represent [O I] in the terrestrial night sky (left) and redshifted in the chromosphere or protostellar disk of TW Hydrae. The code is tuned to ignore known transitions such as this, but the redshift pushed the candidate outside of the regime of exclusion.

Figure 12 shows candidate laser emission our algorithm identified in a well-known T Tauri star, TW Hydrae. The emission is a well-documented forbidden line, [O I] 5577 Å, due to transition in the heated gas in the chromosphere or protostellar disk around TW Hydrae, redshifted slightly (Gorti 2011). In addition, un-redshifted [O I] 5577 A is seen 0.7 Å shortward (at left in bottom panel). But it illuminates the entire slit in accordance with its emission over the whole sky. The barycentric correction for the time of observation (June 23, 2012 beginning of the night) is 23.62 km s$^{-1}$. Combined with the TW Hydra's radial velocity relative to the Sun's proper motion of 13.4 km s$^{-1}$, the total radial velocity the expected topocentric radial velocity is predicted to be then of approximately 37.0 km s$^{-1}$, is consistent with the redshift of +0.7 Å seen in the reduced spectrum. Thus, the laser detection pipeline simply found a known, albeit redshifted, emission line from this T Tauri star.

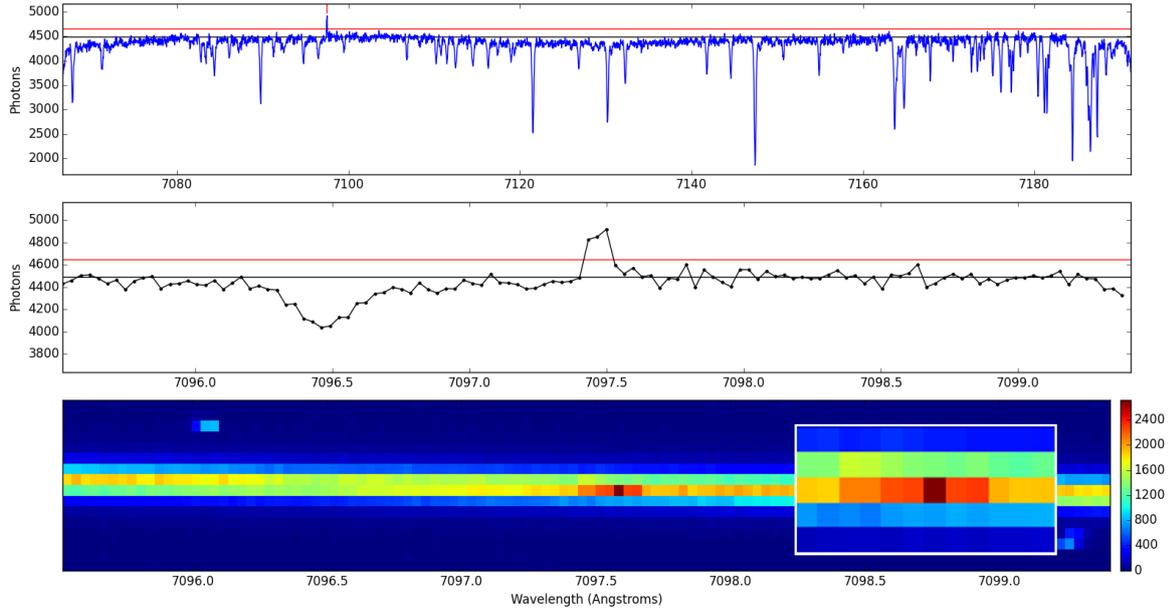

Fig 13. Diagnostic plots for the laser candidate in the spectrum of WASP-1 with the same format as Fig 4. The candidate laser feature is narrower than the instrumental profile in the wavelength direction and narrower than the spatial profile of the star, both discrepancies indicating that the feature did not originate from light passing through the spectrometer. A zoom-in on the central 12x5 pixels is superimposed on the raw CCD image to emphasize the lack of a noticeable peak in all but the central row. See Fig 14.

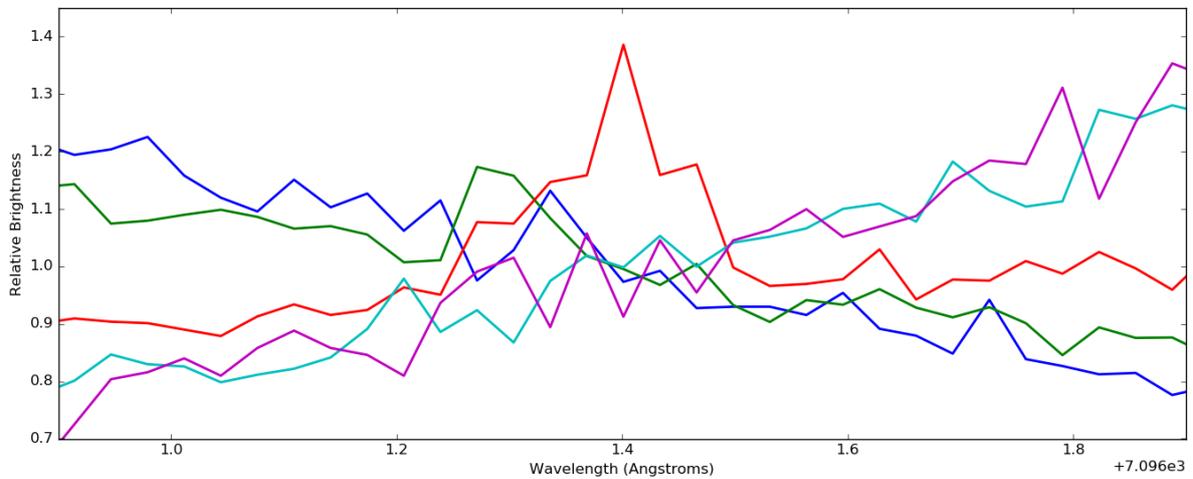

Fig 14. Comparison of the wavelength profiles of 6 rows in the raw image in Fig 13 to show the lack of a peak in all but the central (cyan) row. Pixel values have been normalized by dividing the mean pixel value of the row segment displayed. If there were any light that had passed through the spectrometer in order to produce the peak seen in the central, cyan row, we should see some peak in the adjacent orders. The full-width-half-max of the spatial point-spread function is close to 2.5 pixels for a night of good (i.e. low) atmospheric seeing, as was the night on which this spectrum was taken. The eye can pick out no peaks in any of the adjacent orders.

Finally, Figure 13 shows a small peak of counts that to the untrained eye appears sufficiently narrow as to be consistent with the sort of signal for which we are looking. However, we can see that the counts only appear in the central row of the stellar ridge. That is to say, there is no apparent smearing in the spatial direction. Figure 14 shows the row containing the feature of interest (cyan) plotted with the three rows above and three below in the spatial direction, each row having its mean value normalized to unity. None of the five rows plotted save the stellar ridge show any signs of elevated counts, as they should with the brightness of the detected feature. The increasing and decreasing trends come from the image of the spectral order on the CCD not being perfectly straight.

After this analysis, no potential laser candidates remained from the results of executing the pipeline on 68,636 spectra of 5600 different stars (and a few other astronomical objects). Each prospective laser emission line candidate that was identified by our automated search algorithm had a compelling non-laser explanation upon examination and analysis of the raw CCD image. *Thus we found no convincing laser emission in any of the 68,636 spectra of 5600 stars.*

## 5. Flux Limits of Detectable Lasers
### 5.1 Spectrophotometric Calibration

It is important to quantify the laser emission nondetections and the associated detection thresholds in a way that permits translation into the quantitative implications for the prevalence and characteristics of such lasers. We noted that nearly all of our spectra fell into four predominant stellar spectral types, F, G, K, and M main sequence stars, comprising over 95% of our targets.

We determined our detection thresholds as follows. We used the results of the injection and recovery of synthetic laser emission lines that were sampled at 48 representative wavelengths in spectra of four representative spectral types, FGKM, as described in Section 3.4. The blind detection rate of the mock laser lines yielded the detection thresholds necessary for an 80% detection probability of laser emission in units of the star's spectral continuum. We endeavored to place these stellar continua on a scale of absolute flux densities to yield the laser detection threshold intensities in terms of energy emitted per unit time, per unit wavelength interval, into a specified solid angle.

For each of the four spectral types, we identified one representative star that had been studied carefully with absolute spectrophotometry by Stone (1996). Stone used large-slit spectroscopy and absolute spectrophotometric standard stars to measure the apparent magnitudes at 59 wavelengths from 4167 to 8380 Å. We converted these flux densities to absolute flux densities in units of W m$^{-2}$ Å$^{-1}$ according to the prescription in Stone (1996). We searched for a star studied by Stone in each spectral class that had both a measured accurate parallax and broadband photometry in the U and V bands. The U band photometry offered an approximate measure of the flux densities at 3600 Å, shortward of the shortest wavelength, 4167 Å, provided by Stone (1996). We converted from U-band magnitude to flux density using the following expression:

$$Flux_U = 4.22 \times 10^{-12} \times 10^{(-U_{mag}/2.5)} \text{ W m}^{-2}\text{ Å}^{-1}. \quad (1)$$

The four representative stars we identified in Stone (1996) are listed in Table 1, giving the name, spectral type, V-band magnitude, and U-band magnitude from SIMBAD. For each of those stars, the Stone spectrophotometry was converted to flux densities in units of W m$^{-2}$ Å$^{-1}$ and spline-interpolated onto the wavelengths at which we had previously calculated the detection thresholds in continuum units, namely the central wavelengths of the HIRES spectral orders. Multiplying those continuum-normalized detection thresholds by the absolute flux densities yielded the absolute flux densities of the laser emission that would have been detectable. A simplifying assumption was made in not treating the integrated number of photons over the entire Gaussian PSF. Rather, the total flux was approximated by the product of the peak pixel value quoted in Figure 3 and the instrumental profile width (equal to the FWHM of the Gaussian model) of approximately 4.2 pixels.

Table 1: Spectrophotometric Standard Stars

| Star | Spectral Type | Vmag | Umag |
| --- | --- | --- | --- |
| HD 24538 | F5V | 9.58 | 10.02 |
| 16 Cyg B | G3V | 6.2 | 7.07 |
| G104-337 | K2V | 11.22 | 12.8 |
| GJ 1142A | M3V | 12.56 | 15.25 |

References: (1) SIMBAD (Wenger et al. 2000), (2) Stone 1996

We then scaled the flux densities for each of the four spectral types in Stone (1996) to what they would be for an apparent V-band magnitude that is roughly representative of the V-band magnitude of the majority of the target stars within each spectral type, namely V=7 for F stars, V=8 for G stars, V=8 for K stars, and V=10 for M stars. (The stars from planet transit surveys are considerably fainter, V = 9 to 13, and those in the *Kepler* field are fainter still). These adopted nominal magnitudes are not important as the detection threshold flux densities can always be scaled to any specific magnitude for a host star of the laser emission. We felt it was useful to present thresholds for the flux densities of laser emission that pertain to the majority of our target stars. For any particular star, the detection threshold can be easily obtained by scaling the fluxes at a fiducial magnitude such as V-band, tantamount to accounting for the ratio of distances. For example, a G star having V magnitude of 13 will have flux densities that are 1/100 that of the V=8 reference G star, at all wavelengths.

## 5.2 Laser Flux Density Thresholds

Figure 15 shows the resulting stellar flux densities for the nominal magnitudes of the four spectral types in the four panels. Each panel also shows the detection threshold for laser emission in units of W m$^{-2}$ Å$^{-1}$. The plots show the detection thresholds for spectra obtained with iodine (in red) and without iodine (in blue) in the HIRES spectra. The iodine absorption lines contaminate the spectrum with lines of typical depth 20% so that a laser line is more difficult to detect, both because it may reside deep within an iodine absorption line diminishing its contrast with the neighboring stellar spectrum and because the choppy spectrum renders the continuum less well defined, forcing laser emission to be brighter to stand out.

The typical laser detection thresholds in flux density are roughly $10^{-15}$ W m$^{-2}$ Å$^{-1}$, comparable to the stellar flux densities, but there is a strong dependence on spectral type of the host star and on wavelength. For touchstone G3V stars, the near UV and blue wavelengths are chopped up by myriad atomic neutral metal absorption lines, making the detection of laser emission more difficult for the same reason that iodine lines compromise laser line detection. Longward of 5000 Å, the spectra of G stars become systematically smoother, lowering the detection thresholds of laser lines because they stand out more easily against the smooth continuum. This is partially due to the criterion set in our algorithm that demands three consecutive pixels be above our threshold. Higher effective local RMS noise from the stellar flux distribution itself increases the likelihood that one pixel will sit low enough to prevent the laser line from exceeding our threshold in that pixel. The K-type stars exhibit a similar behavior with wavelength in their detectability of laser lines. For the M3V stars, their fluxes are so low in the near UV and blue that a laser line need have only comparable flux to stand out against the stellar flux. The resulting laser detection thresholds are typically lower than $0.2 \times 10^{-15}$ W m$^{-2}$ Å$^{-1}$. Toward longer wavelengths, the M dwarfs have higher flux densities, forcing the laser flux density to be correspondingly higher to compete, reaching $0.7 \times 10^{-15}$ W m$^{-2}$ Å$^{-1}$ at 7000 Å. For the F5V stars, their higher effective temperatures raise the continuum flux in the UV and blue, forcing the laser emission flux density to be quite high for detection, as much as $1 \times 10^{-15}$ W m$^{-2}$ Å$^{-1}$. Toward longer wavelengths, the flux density of F stars diminishes, bringing the detection threshold down to levels below $1 \times 10^{-15}$ W m$^{-2}$ Å$^{-1}$.

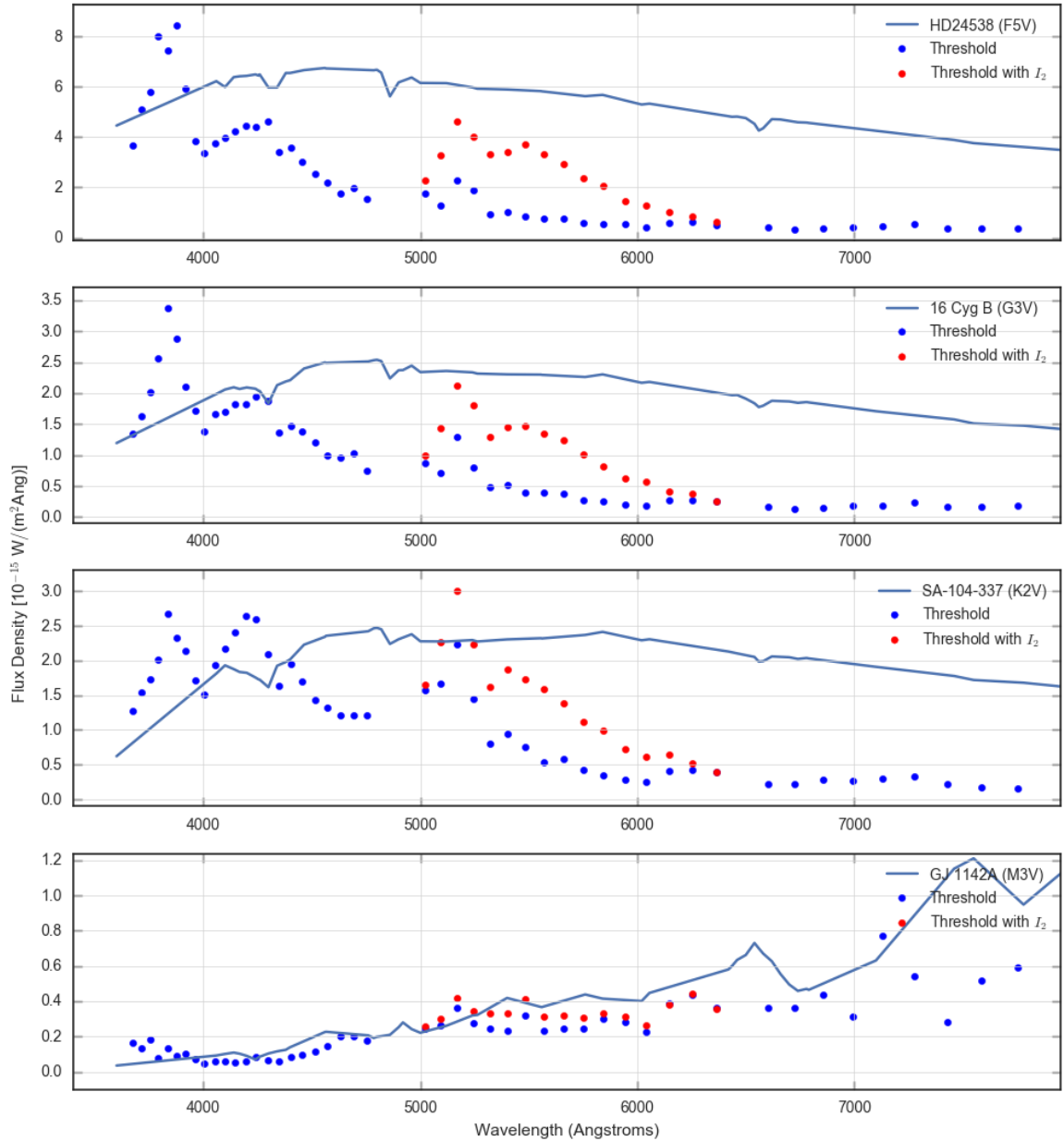

Figure 15. Flux density versus wavelength for F5, G3, K2 main sequence, and M2 dwarf stars (blue) with typical corresponding magnitudes, Vmag = 7, 8, 8, and 10, and the associated flux density detection thresholds for laser emission superimposed on such a spectrum. Detection thresholds for spectra without iodine contamination (blue points) and with iodine (red) vary significantly by wavelength and among spectral types. F star thresholds vary between $10^{-14}$ W m$^{-2}$ Å$^{-1}$ in the near UV (at left) to less than $1\times10^{-15}$ W m$^{-2}$ Å$^{-1}$ at 7800 Å. The G thresholds vary over the same wavelength range from $3.4\times10^{-15}$ W m$^{-2}$ Å$^{-1}$ to $0.2\times10^{-15}$ W m$^{-2}$ Å$^{-1}$. The K stars exhibit slightly higher thresholds in the blue, and range from $2.6\times10^{-15}$ W m$^{-2}$ Å$^{-1}$ to $0.2\times10^{-15}$ W m$^{-2}$ Å$^{-1}$. Finally, the M stars show detection thresholds significantly lower due to the inherent low luminosity of such stars, ranging from $0.1\times10^{-15}$ W m$^{-2}$ Å$^{-1}$ to $0.8\times10^{-15}$ W m$^{-2}$ Å$^{-1}$. The variation in the detection threshold for laser emission was computed by trials of injection and recovery of synthetic laser emission lines inserted directly into our spectra and analyzed blindly to determine if they would have been detected.

## 5.3 Laser Power Thresholds

We now compute benchmark laser power required for detection by adopting a minimum wavelength interval for the laser emission "line" and adopting a nominal solid angle for the laser beam. The detection algorithm for laser emission requires that the emission have a width at least as wide as the FWHM of the instrumental profile of the HIRES spectrometer, which is 4.3 pixels with variations of only a few tenths of a pixel over the entire echelle format. We used the wavelength scale to convert that 4.3-pixel width to a width in wavelength, $\Delta\lambda$, which varies from 0.01 to 0.04 Å due to the changing dispersion throughout the echelle format. At 5000 Å, the FWHM of the instrumental profile is $\Delta\lambda = 0.022$ Å.

We adopted a fiducial solid angle for the hypothetical laser beams, enabling future scaling to any solid angle. Obviously one has no idea what solid angles might be appropriate for laser emission from extraterrestrial civilizations or any other sources. We adopt a touchstone optical device having diffraction-limited performance comparable to the largest optics on Earth, i.e., diffraction-limited mirrors of 10-meter diameter, either in space or on the ground corrected for atmospheric refraction. Adhering to this touchstone optic, we consider an extraterrestrial laser launcher characterized by diffraction-limited, 10-meter diameter optics, emitting a beam having its first interferometric null at an angle, theta = 1.2 $\lambda$/D, from the optical axis, where D is the effective diameter of the launching optic.

The conical shape of the laser beam intercepts the Earth with a circular footprint that has an area, $\pi r^2$, where r is given by r = $\theta$ d, where d is the distance from the laser to the Earth. Thus, the footprint area of the laser beam at the Earth is given by, $A = \pi (1.2 \lambda d/D)^2$. The footprint area, at Earth, of the laser thus increases as the square of both the wavelength and the distance to the laser. We multiply the flux density laser detection thresholds by the instrumental profile width, $\Delta\lambda$, given above, and the laser footprint area, A, to yield the power of the laser that is just detectable, in units of Watts, for the touchstone diffraction-limited 10-meter laser.

We note that the distance to the star, d, matters little in computing these detection thresholds. The reason is that we chose exposure times for the stellar spectra that yielded a nearly consistent number of photons per pixel, thereby compensating for distance. That is, no matter the distance to the star, we used exposure times sufficiently long enough to reach similar signal-to-noise ratios in the stellar spectra, thereby achieving similar detection thresholds, relative to the stellar continuum, for any laser emission lines.

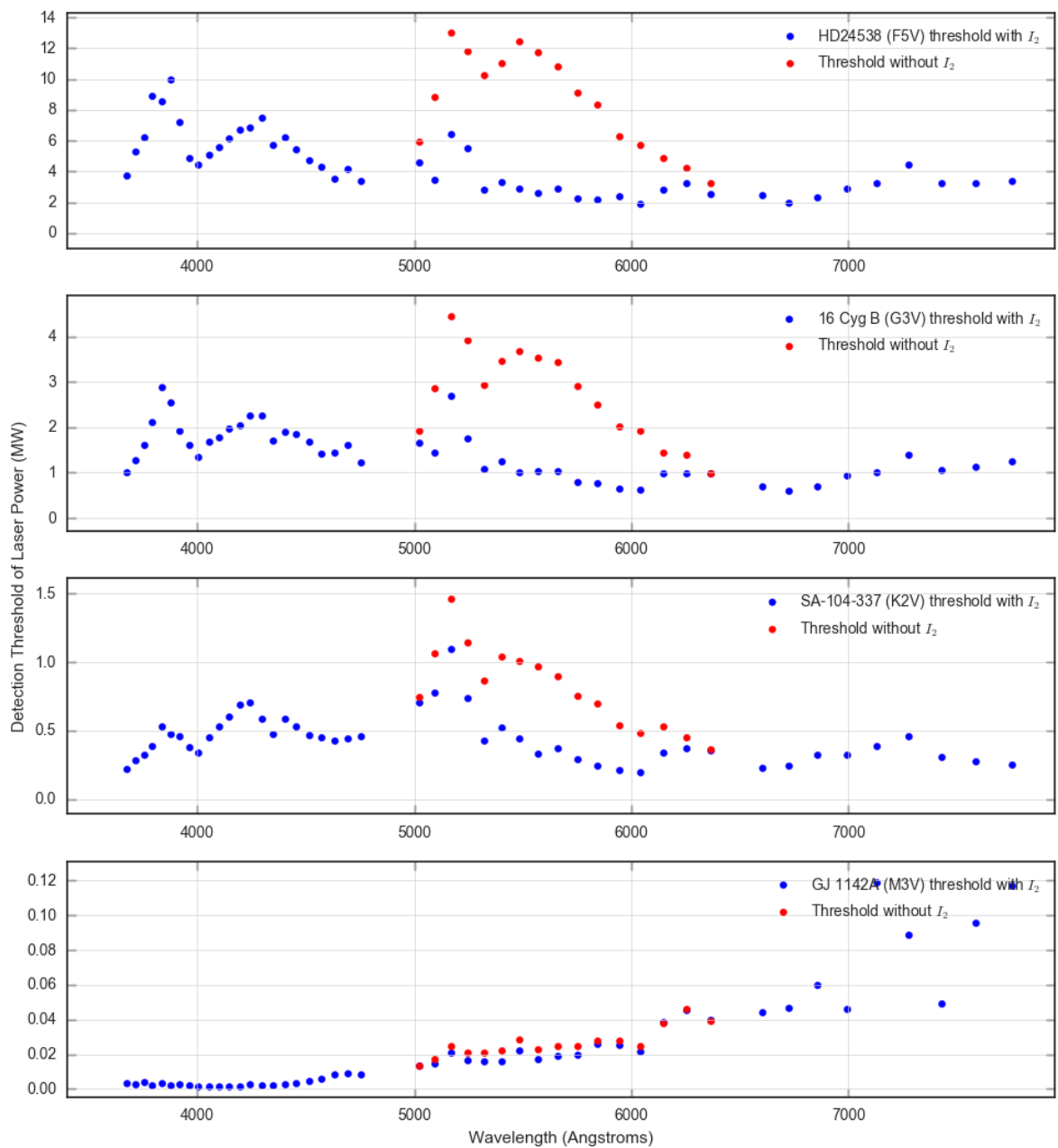

Figure 16. Detection threshold versus wavelength for the power from just-detectable diffraction-limited lasers launched from 10-meter optics located near F5, G3, K2, and M2 main sequence stars. Spectra taken without the iodine cell are shown in blue and those with the iodine cell are shown in red. The laser power required for detection varies between several kilowatts to 13 megawatts, with the variations caused by the competing spectral energy distribution of the stars and the choppiness of the spectra from absorption lines. Of special note are the thresholds for the faint M dwarfs, for which the threshold laser power for detection ranges from 1 to 120 kilowatts, depending on wavelength and only weakly on the presence of iodine lines. For other stars of the same spectral type, thresholds will be similar to within approximately 50%.

The resulting detection thresholds for the average power (in megawatts) of the benchmark diffraction-limited lasers are shown in Figure 16, corresponding to the detection of unresolved lasers spatially unresolved from host stars having representative spectral types mentioned previously, F5, G3, K2, and M3. Each figure shows detection thresholds of laser power vs wavelength for each of the spectral types. The overarching result is that diffraction-limited lasers would be detected having average power over our several-minute exposures of a few kilowatts to 10 megawatts, depending on spectral type and wavelength, as follows.

For F5 stars, Figure 16 shows that a laser must have an average power between 2 and 13 megawatts in the UV and blue regions of the spectrum to be detectable in our spectra. Longward of 5000 Å, the detection thresholds are between 2 and 6.5 megawatts, lower than in the UV and blue because of the paucity of absorption lines, yielding a smoother spectral continuum that allows laser emission to stand out. For spectra of F5 stars obtained with the iodine cell in the spectrometer, the wavelength region between 4950 and 6300 Å is chopped by iodine absorption lines, hindering detection of laser emission, causing higher detection thresholds, up to 13 megawatts. All of these thresholds pertain to a diffraction-limited laser launcher of 10-meter diameter. The thresholds can be easily scaled to lasers having wider or narrower beams, with threshold scaling as the square of the diameter of the diffraction-limited optics.

For G3 stars, Figure 16 shows that a laser must have average power of 1 to 3 megawatts in the UV and blue regions of the spectrum to be detected in our spectra. Longward of 5000 Å, the detection thresholds are between 0.6 and 2.7 megawatts, lower than in the UV and blue because of the smoother stellar continuum spectrum. For spectra of G3 stars obtained with the iodine cell the detection thresholds are raised to between 1.0 and 4.5 megawatts.

For K2 main sequence stars, Figure 16 shows the laser power detection thresholds to be 0.2 to 1.1 megawatts throughout the entire wavelength region, 3600 to 8000 Å. The roughly uniform density of absorption lines and broad spectral energy distribution throughout those wavelengths leaves the detectability of laser emission roughly comparable at all of those wavelengths. For spectra with iodine lines, the detection thresholds are raised by a factor of roughly 2 between 4950 and 6300 Å.

For M2 dwarfs, Figure 16 shows the laser power detection thresholds for spectra obtained without the iodine cell (yellow) to rise from only 2 kilowatts at 3600 Å to 120 kilowatts at 8000 Å. For spectra with iodine lines, the detection threshold is only raised by roughly 20% as the stellar spectra are already littered by molecular lines of their own, notably from oxides and hydrides in the stellar atmosphere. It is noteworthy that the low optical luminosity of M dwarfs allows laser emission from any unresolved orbiting planets to be detectable even if they have only kilowatt power levels.

6. Implications for Pulsed Optical SETI

The detection thresholds in this survey can be translated to limits on searches for pulses of optical emission from extraterrestrial sources, such as the searches by Howard et al. (2004), Hanna et al. (2009), and Wright et al. (2014b). We may characterize optical pulses by the time separation between pulses, $\Delta t_{sep}$, and the number of photons per unit area per pulse hitting Earth's atmosphere, $N_{pulse}$. During a telescope "exposure" of length $t_{exp}$, a telescope of aperture area, $A$, will capture a train of optical pulses containing a total number of photons given by:

$$N_{exp} = (t_{exp}/\Delta_{sep}) * N_{pulse} * A . \quad (2)$$

The present spectroscopic survey would capture such a train of pulses, albeit unresolved temporally, if the emission were confined in wavelength to widths less than a few Å. Figures 15 and 16 permit the determination of the total number of photons within a laser pulse train that would have been detected here. Our stellar spectra typically have ~40,000 photons per pixel (reduced), implying that the 4.3 pixel PSF FWHM contains ~170,000 stellar photons. Our detection thresholds are roughly 30% of the stellar flux (see Figure 15), implying a threshold number of photons within an emission line of ~50,000. Any train of optical pulses containing more than 50,000 photons during our typical 10-minute telescope exposures would have been detected here.

As a benchmark example, we may imagine photon pulses that arrive every second ($\Delta t_{sep} = 1$ sec). We adopt a typical telescope exposure time of $t_{exp} = 10$ min, thereby capturing 600 pulses. The area of the aperture of the Keck telescope is $A$=76 m$^2$ (including the secondary mirror obstruction). We use the detection threshold for the total number of photons of Nphot$_{exp}$ = 50,000 photons. Thus only 83 photons per pulse are required for detection with our Keck telescope system, i.e., 1.1 photons per pulse per square meter.

Two other examples offer benchmarks. A pulse separation of 60 seconds requires only 66 photons per pulse per square meter for a detection here. A pulse separation of a millisecond requires only 0.001 photons per pulse per square meter. (We have ignored here explicit consideration of the efficiency of the Keck telescope and HIRES spectrometer as the threshold number of photons for detection, 50000, represents the photons detected at the CCD detector, thus including the efficiency.) Thus, for any pulse separation less than 1 second, the present system would have detected all such pulses containing at least 1 photon per square meter per pulse. For pulse separations greater than 1 second, detectable pulses must have a number of photons per square meter equal to the duration measured in seconds (because the threshold number of photons, 50000, happens to equal the product of the typical exposure time, 600 secs, and the area of the Keck telescope, 76 m$^2$).

The reason this spectroscopic system is capable of detecting photon pulses so well is that many pulses are contained in a train of pulses lasting 10 minutes and the Keck telescope has a considerable collecting area. A remarkably small number of photons per pulse per square meter is required for the Keck telescope to accumulate 50,000 photons during the typical 10 minute exposures. One may imagine a pulse train designed to contain a message, with pulse separations of microseconds or milliseconds. The present Keck survey would detect such a train of pulses if each pulse contained just 1 photon per square meter (or even less). Thus this spectroscopic optical SETI program covers the pulsed optical SETI domain carried out with 1-meter class telescopes for all pulses arriving more frequently than one per second. For slower arrival rate of pulses, this survey also covers the domain of photons per pulse equal to the pulse separation in seconds. The nondetection of laser emission here shows that none of the 5600 stars had trains of optical pulses detectable by 1-meter class telescopes for pulse separations less than 1 second and for longer pulse separations given proportionately more photons per pulse.

Note that our method is not very sensitive to changes in pulse width and duty cycle, as long as the total integrated power is high enough. In contrast, for pulsed optical SETI searches the pulse width does matter, as the technique leans on the relative paucity of photons from the star during the nanosecond sampling or during the pulse duration. Pulses of longer pulse width may still be detected by the photon counting detectors used in Wright et al. (2014b). Photon flux must be approximately that coming from the star, or more, at some point during the pulse in order to be detected. On the other hand, searches for pulses in the near IR remain utterly unique and unexplored (Maire et al. 2014). Also, pulses that span a range of wavelengths more than ~10 Angstroms are not easily detected by the spectroscopic method, as the spectroscopic resolution provides diminishing contrast with the rest of the spectrum. Our spectroscopic approach offers a domain of advantages because it suppresses the contamination from the starlight by spectroscopically isolating the laser emission line from 99.99% of the starlight at all other wavelengths. Also, it piggybacks on existing spectra obtained with 10-meter class telescopes, and yet requires no telescope time.

Future all-sky optical SETI programs should consider quantitatively the increase in parameter space they will sample relative to the parameter space ruled out here. Apparently next generation all-sky optical SETI must sample an order of magnitude more stars (or sky) at least. The detection thresholds here of 1 photon per pulse per square meter will be superseded by the upcoming Breakthrough-Listen program, for which detailed modeling of the stellar spectrum will reduce the threshold to ~5% of the continuum intensity, dropping the laser power required for detection by another factor of 10.

7. *Breakthrough Listen*: A New Spectroscopic Laser Search

During three years, through 2018, the *Breakthrough Listen* initiative will use the 2.4-meter "Automated Planet Finder" (APF) optical telescope 36 nights per year with its "Levy Spectrometer" to search for laser emission (Isaacson et al. 2017), similar to the program described here with the Keck Observatory. The Levy spectrometer acquires spectra spanning wavelengths 374 to 980 nm with a resolution, R = 95,000 when using a nominal 1 arcsec slit, 50% finer resolution than that of the Keck-HIRES spectrometer. The APF-Levy instrument acquires spectra only 6 times more slowly than Keck-HIRES due to the 20x smaller telescope aperture but 3x faster spectrometer design (Vogt et al. 2014, Radovan et al. 2014). *Breakthrough Listen* will acquire spectra with a signal-to-noise ratio of typically 100:1 (1% noise per pixel) to achieve detection thresholds for laser lines quite similar to those reported here for Keck-HIRES.

The optical target list within the *Breakthrough Listen* program contains 1709 stars of all spectral types, OBAFGKM, and both main sequence stars and giants. The *Breakthrough Listen* program will concentrate on stars of spectral type O-, B-, A-, F-, and late M-type, to complement the survey reported here (Isaacson et al. 2017). It will also continue to observe stars of G- and K- type, both in common with and supplementary to those GK stars observed here. The targets will be closely similar to those of the *Breakthrough Listen* Green Bank radio program, notably a complete sample of all 43 stars within 5 pc northward of declination -20 deg, 1000 nearby stars with a diverse set of spectral types, and the centers of 100 nearby galaxies including 30 spirals, 30 ellipticals and 40 dwarf spheroidals and irregulars. *Breakthrough Listen* will take spectra with the APF with the assigned telescope time, but will also use the approximately 3000 archived spectra taken with the APF-Levy spectrometer. In the interest of supporting panchromatic SETI, the NIROSETI program (Wright et al. 2014) will observe the same targets observed in the optical under the *Listen* program.

The O, B, and A target stars are particularly intriguing because the additional UV flux compared to G and K stars might augment the rate of mutations, allowing such short lived host stars (lifetimes under 1 billion years) to accelerate evolution toward intelligent life. The lowest mass M-type stars are also intriguing, as the five small planets around Trappist-1 and the fundamental results of Dressing et al. (2015) show that Earth-size planets may be common around the smallest stars and even brown dwarfs.

The signal-to-noise of 100 of the APF-Levy spectra will afford laser detection thresholds relative to the continuous flux of the star comparable to that in this current work, i.e. corresponding to laser power thresholds of 10 kilowatts to 10 megawatts. The *Breakthrough Listen* APF-Levy optical SETI program will complete a survey of all common types of stars in the Milky Way Galaxy, providing a statistically deep sampling of laser emission from technological civilizations, especially those of enough advancement to harbor a multitude of laser beams or to be purposely attempting to contact us.

## 8. Discussion and Conclusions

We have searched the spectra of 5600 stars for optical laser emission using the Keck 1 telescope and its high resolution spectrometer, HIRES. We examined multiple spectra of most of the stars, obtained during a time span of many years, for a total of 68,336 spectra. The entrance slit of the spectrometer gathered the light from within 0.5 x 1.0 arcseconds from the star, corresponding to the entire region within a few tens of au from the star, depending on distance to the system, most being tens to hundreds of pc away. The habitable zones were thus observed around each star for laser emission, and usually 10x farther out as well, to include most of the planets in each planetary system. The spectra have high resolution (lambda / delta lambda = 60,000) and high signal-to-noise ratio (over 100 per pixel), making these spectra well suited for detecting laser emission having an intensity that is more than a fraction of the stellar intensity at each wavelength.

We computed the detection thresholds by injecting thousands of synthetic laser lines into the actual spectra of all types and wavelengths observed here and blindly executing our search algorithm to determine the probability of detecting them. We calibrated those detection thresholds in energy flux units using absolute spectrophotometry of standard stars. The final detection thresholds correspond to lasers having continuous power in the range of a few kW to 13 MW, depending on the spectral type of the host star. As a benchmark, we computed the detection thresholds of laser power corresponding to "Keck-to-Keck" diffraction-limited laser beams from a 10-meter diameter emitter, yielding laser opening angles of roughly 0.01 arcsecond, depending on wavelength. Both such laser power and opening angles are achievable by contemporary human technology.

Indeed, it is useful to define a standard reference beam geometry to communicate the detectable specific intensity from optical, IR, and UV lasers of unknown beam size. Radio SETI often adopts "equivalent isotropic radiated power" (EIRP) as a metric of transmission intensity, as the actual solid angle of any prospective extraterrestrial radio transmitter remains unknown. Optical laser beams of mere meter-size optics have intrinsically tiny cone angles, typically less than 1 arcsec, corresponding to solid angles less than $10^{-10}$ sr. We adopt a benchmark optical laser launcher to form an analogous geometrical metric for the sought optical lasers. A diffraction-limited 10-meter diameter optic constitutes a reference fiducial laser beam launcher that is consistent with contemporary technology, as adopted by others (e.g. Ekers et al 2002, Howard et al. 2004, Hanna et al. 2009). The emitted solid angle (and the illuminated footprint area at the receiver) of such a beam launcher is proportional to the square of the wavelength, $\lambda$, of the laser light and inversely as the square of the diameter, D, of the emitting optic, $\Omega = (\lambda/D)^2$, allowing subsequent scaling to any envisioned wavelength and launch optic diameter. This reference beam launcher forms an analogous metric for the received intensity, namely the radiated power from a diffraction-limited ten-meter aperture or "equivalent ten-meter radiated power", ETRP. We report the laser power detection thresholds for the 5600 stars in this present survey in terms of this reference ETRP in Section 5 and Figure 16, yielding detectable laser power spanning the range 3 kW to 13 MW, depending on wavelength and star spectral type.

We may imagine that beings more technologically advanced than humanity would be capable of constructing ETRP laser launchers with power levels at least as high as those detectable here, for any of the 5600 star systems we surveyed. Our blind search algorithm identified several dozen candidate laser emission lines. But our initial eyeball vetting rejected obvious false positives caused by unusual cosmic ray tracks, instrumental flaws, or Earth-borne atmospheric emission, leaving 12 surviving candidate laser lines. Subsequent detailed examination of those 12 candidates compelled us to reject a laser origin for each, as described in Section 4. *Thus, we found no compelling evidence for extraterrestrial laser emission among any of our 5600 stars at power levels of 3 kW to 13 MW.*

It is incumbent on searches for extraterrestrial technology to translate non-detections in terms of prospective entities that could have produced such signals, whether biological or machine. Unfortunately, there are no widely accepted models of the Milky Way Galaxy that predict the properties of extraterrestrial technological entities, such as their prevalence, capabilities, or intentions, against which to compare the present non-detections of megawatt-level laser emission from 5600 stars. Similarly, authoritative models do not exist for their distribution spatially throughout the Galaxy, their typical lifetimes, nor their advancement in technology, especially optical beaming. Remarkably, science fiction and philosophical soothsayers offer the closest facsimiles to Galactic models of technological beings (e.g., Benford 1984, Brin 2012ab, Schneider 2016). We are left to conjure such models with little constraint.

If any of the 5600 stars we surveyed contained optical lasers of megawatt power (ETRP) located within a few au of the star and pointed toward Earth at the time of our observations, they would have been detected. To be detected the lasers would have to be aimed at where the Earth will be located upon arrival of the beam, given the projected future position of the Earth and the light-travel time. Such position and velocity data are already available to humanity for other stars and their known transiting planets if any, notably with the data from the Hipparcos, Gaia, and *Kepler* telescopes. Such informed laser systems around any of our 5600 target stars allow technological civilizations to reveal themselves to us, either purposefully or inadvertently. For inadvertently oriented laser beams to strike the Earth would require one of two conditions: either fortuitous pointing of a narrow laser beam toward the future location of Earth or a fireworks-like display of trillions of laser beams pointing away from the host, filling a substantial fraction of their entire sky to serendipitously intercept Earth. Alternatively, an advanced civilization can simply point a MW laser at Earth.

As Earth-size planets exist in the habitable zones around 20 to 40% of nearby stars (Petigura et al. 2014, Dressing and Charbonneau 2015), many of the 5600 stars we surveyed have such planets. This implies that our laser search sampled ~2000 Earth-size planets in their habitable zones. Those 2000 lukewarm Earth-size planets could, in principle, be hosting technological civilizations, if biology and evolution happen elsewhere.

Our 5600 non-detections of narrow-band light emission shows that none of those ~2000 lukewarm Earth-size planets host a MW laser beamed toward us. One simple model of the Galaxy has life commonly and quickly occurring on Earth-size planets in their habitable zones. One possible model of exobiology is that microbial life commonly forms and evolves inexorably toward intelligence that lasts millions of years, with a corresponding development of technology. Indeed, almost all science fiction is based on this model. However, the current nondetection of laser signals suggests limits on this model. If even a small fraction of the ~2000 lukewarm Earth-size planets surveyed here had technological civilizations beaming megawatt-level lasers (ETRP) toward Earth, we would have detected them. As a touchstone, if technological beings emerge on just 1% of such Earth-size planets, our survey sampled laser emission from roughly 20 such civilizations. Indeed, a common suggestion is that advanced civilizations would be thousands or millions of years more advanced than humans. If so, they would likely know of our existence and our technological capabilities (and limitations). They could extend a laser finger, hoping to touch the youthful humans who only this century gained the ability to reach out with light beams of their own. Unfortunately, we saw no such effort at communication. We rule out models of the Milky Way in which over 0.1% of warm, Earth-size planets harbor technologies that, intentionally or not, are beaming optical lasers toward us. We may begin to wonder if arguments along the lines of the so-called Fermi paradox have some merit (e.g. Zuckerman 2002). This upper limit sets a requirement for next-generation optical SETI searches, that they survey an order of magnitude more stars, either targeted or blind wide-field, or achieve much lower detection thresholds than megawatt power lasers.

## Acknowledgments

We thank the Templeton Foundation and the Breakthrough Prize Foundation for support of this research. We are grateful to the Keck Observatory and its staff, especially Scott Dahm, Greg Doppman, Hien Tran, Grant Hill, and Barbara Schaeffer for support of the Keck 1 telescope, HIRES spectrometer, and Bob Kibrick and Greg Wirth for remote observing technology. We are grateful to the University of California, NASA, Caltech, and the University of Hawai'i for their stewardship and support of the Keck Observatory, and also for their support of the exoplanet research that allowed this additional use of the same spectra. We are grateful to the Keck Observatory Archive for their stewardship and documentation of all spectra taken with the HIRES spectrometer (and other Keck instruments). We thank the many observers who contributed to the spectra reported here, especially leading scientists, Debra Fischer, Jason Wright, John Johnson, Kathryn M.G. Peek, Paul Butler, Steve Vogt, Howard Isaacson, Erik Petigura, Lauren Weiss, and Lea Hirsch. We thank Gloria and Ken Levy for support of graduate student Lauren Weiss. We gratefully thank Dan Werthimer, Andrew Siemion, Paul Horowitz, Jill Tarter, Seth Shostak, Frank Drake, Jason Wright, John Gertz, and Andrew Fraknoi for many conversations about the search for intelligent life in the universe. This work made use of the SIMBAD database (operated at CDS, Strasbourg, France) and NASA's Astrophysics Data System Bibliographic Service. We thank the Keck Observatory Archive (KOA) for making all spectra analyzed here publicly available. We also wish to extend special thanks to those people of Hawai'ian ancestry on whose sacred mountain of Maunakea we are privileged to be guests. Without their generous hospitality, this study of our shared, beautiful universe would not be possible.

# References

Abeysekara, A. U., Archambault, S., Archer, A., Benbow, W., Bird, R. et al., 2016. ApJL 818, L33

Alonso, R., Brown, T. M., Torres, G., Latham, D. W., Sozzetti, A., et al., 2004, ApJ 613 L154-L156

Arnold,L.F.A. 2005a, Transit Light-Curve Signatures of Artificial Objects. ApJ, 627, 534

Bakos, G. Á, Hartman, J. D., Torres, G., Béky, B., Latham, D. W. et al., 2012, AJ 144, 1-19

Bakos, G. A., Csubry, Z., Penev, K., Bayliss, D., Jordán et al. 2013, PASP 125, 924

Benford, G. 1984, "Across the Sea of Suns", Timescape Books, ISBN0-671-44668-1

Bodman, E. H. L., Quillen, A., 2016. ApJL 819, L34

Boyajian, T.S., LaCourse, D.M., Rappaport, S.A., et al. 2016, MNRAS, 457, 3988

Breakthrough Prize Foundation 2016 http://breakthroughinitiatives.org/Initiative/3

Brin, D. 2012a, Nature, 486,471

Brin, D. 2012b, "Existence", Tor Books, ISBN 978-0-765-30361-5

Cameron,A.C. 2016, Methods of Detecting Exoplanets, Astrophysics and Space Science Library, V.428, p.89

Cocconi, G., & Morrison, P., 1959, Nature 184, 844-846

Collier C.,A, Pollacco,D., Hellier,C. et al. 2007, IAUS, 253, 29

Covault, C. 2013, Poster at APS April Meeting (Denver), http://meetings.aps.org/Meeting/APR13/Session/S2.2

DARPA, 2015 Press release, http://www.darpa.mil/news-events/2015-0

Drake, F. D. 1961, Physics Today, 14, 40

Drake, F. D., Werthimer, D., Stone, R. P. S., & Wright, S. A. 2010, in Astrobiology Science Conference, Evolution and Life: Surviving Catastrophes and Extremes on Earth and Beyond (Houston: Lunar and Planetary Institute), http://www.lpi.usra.edu/meetings/abscicon 2010/pdf/5211.pdf

Dressing, C. D., Charbonneau, D., 2015, ApJ 807, 45

Dyson,F.J. 1960, Science, 131, 1667

Ekers,R.D. et al. 2002, "SETI 2020: A Roadmap for the Search for Extraterrestrial Intelligence", produced for the SETI Institute by the SETI Science & Technology Working Group, Ronald D. Ekers et al. (editors); SETI Press.

Enoch, B., Anderson, D. R., Barros, S. C. C., Brown, D. J. A., 20, Collier Cameron, A. et al. AJ 142, 3-86

Fischer, D. A., Marcy, G. W., & Spronck, J. F. P. 2014, ApJS, 210, 5

Forgan, D. H. 2013, J. British Interplanet. Soc., 66, 144

Gillon, M. 2014, Acta Astronautica, 94, 629

Groom, D., 2004, Scientific Detectors for Astronomy, Springer, see also snap.lbl.gov/ccdweb/ccdrad_talk_spie02.pdf

Gorti, U., Hollenbach, D., Najita, J., & Pascucci, I. 2011, ApJ, 735, 90

Harp, G. R.; Richards, J.; Shostak, S.; Tarter, J. C.; Vakoch, D. A.; Munson, C., ApJ 825, 155

Hanna,D.S., Ball,J., Covault,C.E. et al. 2009, Astrobiology, 9, 34[1]

Horowitz,P. & Sagan,C. 1993, ApJ, 415, 218

Howard, A. W., Horowitz, P., Wilkinson, D. P., Coldwell, C. M., Groth, E. J., Jarosik, N., et al. 2004, ApJ, 613, 1270

Howard, A., Horowitz, P., Mead, C., Sreetharan, P., Gallichio, J., Howard, S., et al. 2007, Acta Astronautica, 61, 78

Howard, A. W., Marcy, G. W., Bryson, S. T., Jenkins, J. J., Rowe, J. F., Batalha, N. M., et al. 2012, ApJS, 201, 15

Howard, A. W., Marcy, G. W., Fischer, D. A., Isaacson, H., Muirhead, P. S., Henry, G. W., et al. 2014, ApJ, 794, 51

Howard, A. & Fulton, B. J., 2016, Accepted to PASP

Isaacson,H., Siemion,A., Marcy,G.W., et al. 2017, Accepted in PASP

Johansson, S., & Letokhov, V. S. 2004, A&A, 428, 497

Kipping,D.M. & Teachey,A. 2016, Monthly Notices RAS, 459, 1233

Korpela, E. J., Anderson, D. P., Bankay, R., Cobb, J., Howard, A., Lebofsky, M., et al. 2011, Proc. SPIE, 8152, 894066

Lacki, B., ArXiV E-print, Apr 2016, http://adsabs.harvard.edu/abs/2016arXiv160407844L

Leeb, W. R., Poppe, A., Hammel, E., Alves, J., Bruner, M., Meingast, S. 2013, Astrobiology, 13, 521

Lubin, P. 2016, http://arxiv.org/abs/1604.01356

Maire, J., Wright, S., Werthimer, D., Treffers, R. R., Marcy, G. W., Stone, R. P. S., et al. 2014, Proc. SPIE, 2056372

Marcy, G. W., Butler, R. P., Vogt, S. S., Fischer, D. A., Wright, J. T., Johnson, J. A., et al. 2008, Phys. Scr., 130, 014001

Marengo, M., Hulsebus, A., Willis, S., 2015. ApJL 814, L15

McCullough, P. R., Stys, J. E., Valenti, Jeff A., Johns-Krull, C. M., Janes, K. A., et al., 2006, ApJ 648, 1228

Mead, C. C. 2013, Ph.D. thesis, Harvard University

Metzger, B. D., Shen, K. J., Stone, N. C., 2017 ArXIV E-print https://arxiv.org/abs/1612.07332

Morton,T, Bryson,S., Coughlin, et al. 2016, ApJ, 822, 86.

Moutou, C., Deleuil, M., Guillot, T., Baglin, A., Bordé, P., et al., 2013, Icarus Vol. 226, 1625

Petigura,E.A., Howard, A.W., Marcy,G.W. 2013, PNAS, 110, 19273

Petigura, E. A., 2015, Ph.D. thesis, UC Berkeley

Radovan, M. V., Lanclos, K., Holden, B P., Kibrick, R I., Allen, S. L., et al., 2014, Proc SPIE, 9145, 12

Reines, A. & Marcy, G. 2002, PASP, 114, 416

Schneider,S. 2016, "Superintelligent AI and the Postbiological Cosmos Approach", in Lursch,A., What is Life? On Earth and Beyond, Cambridge: Cambridge University Press, in press.

Schuetz, M., Vakoch, D. A., Shostak, S. & Richards, J., 2016. ApJL 825, L5

Siemion, A. P. V., Demorest, P., Korpela, E., Maddelena, R. J., Werthimer, D., Cobb, J., et al. 2013, ApJ, 767, 94

Siemion, A. P. V., Benford, J., Cheng-Jin, J., et al. 2014, in Proc. Advancing Astrophysics with the Square Kilometre Array 116, in press

Stone, R. P. S., 1996, ApJS, 107, 423

Stone, R. P. S., Wright, S. A., Drake, F., Muñoz, M., Treffers, R., & Werthimer, D. 2005, Astrobiology, 5, 604

Tan, P.K. and Kurtsiefer, C. arXiv:1607.05897v1, submitted to Astronomy & Astrophysics.

Tarter, J., 2001, Annual Review of Astronomy and Astrophysics 39.1, 511-548.

Tarter, J., Ackermann, R., Barott, W., Backus, P., Davis, M., Dreher, J., et al. 2011, Acta Astronautica, 68, 340

Tellis, N. & Marcy G., 2015, PASP 127, 952

Townes, C. H., & Schwartz, R. N. 1961, Nature, 192, 348

Vogt, S. S., Allen, S. L., Bigelow, B. C., Brown, B., Cantrall, T., Conrad, A., et al. 1994, Proc SPIE, 2198, 362

Vogt, S. S., Radovan, M., Kibrick, R., Butler, R. P., Alcott, B., et al. 2014, PNAS 126, 938

Walkowicz, L., Howe, A. R., Nayar, R., et al. 2014, in American Astronomical Society Meeting 223 (Washington: AAS), 146.04

Wenger, M., Ochsenbein, F., Egret, D., Dubois, P., Bonnarel, F., et al., 2000. Astron. Astrophys. Suppl. Ser., 143 1 9-22

Werthimer, D., Anderson, D., Bowyer, C. S., Cobb, J., Heien, E., Korpela, E. J., et al. 2001, Proc. SPIE, 4273, 104

Wright, J. T. 2005, PASP, 117, 657

Wright, J. T., Fakhouri, O., Marcy, G. W., Han, E., Feng, Y., Johnson, J. A., et al. 2011, PASP, 123, 412

Wright, J. T.; Cartier, K. M. S.; Zhao, M.; Jontof-Hutter, D; Ford, E B. 2016, ApJ, 816, 17

Wright, J.T., Griffith, R., Sigurdsson, S., Povich, M., & Mullan, B. 2014a, ApJ, 792, 27

Wright, J., Cartier, K., Zhao, M., Jontof-Hutter, D., & Ford, E., 2015, ApJ 816, 17

Wright, S. A., Drake, F., Stone, R. P., Treffers, D., & Werthimer, D. 2001, Proc. SPIE, 4273, 173

Wright, S. A., Larkin, J. E., Moore, A. M., Do, T., Simard, L. et al., 2014, Proc SPIE, 9417, 15

Wright, S. A., Werthimer, D., Treffers, R. R., Maire, J., Marcy, G. W., Stone, R. P. S., et al. 2014b, Proc. SPIE, 9147, 91470 J

Zubrin, R. 1995, ASP Conference Series, V.74 "Progress in the Search for Extraterrestrial life" Ed. G.Seth Shostak

Zuckerman,B 1985, Acta Astronautica, 12, 127

Zuckerman, B. 2002, Mercury, 31, 14, "Why SETI Will Fail"